\documentclass[12pt]{article}

\usepackage{scicite}

\usepackage{times, soul, color, url}
\soulregister\cite7
\soulregister\ref7
\soulregister\pageref7
\DeclareRobustCommand{\hll}[1]{{#1}} 

\usepackage{amsmath}
\usepackage{amsfonts}
\usepackage{amssymb}
\usepackage{graphicx}
\usepackage{titlesec,upgreek,adjustbox}
\newcommand{\mum}{\,\mathrm{\upmu m}}
\usepackage{color}

\usepackage{enumitem}
\usepackage{caption}

\usepackage{multirow}
\usepackage[table,xcdraw]{xcolor}

\usepackage{indentfirst}

\usepackage{soul}

\newenvironment{sciabstract}{%
\begin{quote} \bf}
{\end{quote}}

\title{Wide-Field Multiphoton Imaging Through Scattering Media Without Correction}

\author
{Adri\`{a} Escobet-Montalb\'{a}n,${}^{1}$ Roman Spesyvtsev,${}^{1,\dagger}$ Mingzhou Chen,${}^{1}$ \\
Wardiya Afshar Saber,${}^{2}$ Melissa Andrews,${}^{3}$ C. Simon Herrington,${}^{4}$ \\
Michael Mazilu,${}^{1}$ Kishan Dholakia,${}^{1,\ast}$\\
\\
\normalsize{${}^{1}$SUPA, School of Physics and Astronomy, University of St Andrews,}\\
\normalsize{North Haugh, St Andrews, KY16 9SS, UK.}\\
\\
\normalsize{${}^{2}$School of Medicine, University of St Andrews, North Haugh, St Andrews, KY16 9FT, UK.}\\
\\
\normalsize{${}^{3}$Biological Sciences, University of Southampton, University Road, Southampton, SO17 1BJ, UK.}\\
\\
\normalsize{${}^{4}$CRUK Edinburgh Centre, Institute of Genetics and Molecular Medicine, }\\
\normalsize{The University of Edinburgh, Crewe Road South, Edinburgh, EH4 2XR, UK.}\\
\\
\normalsize{${}^{\dagger}$Present address: SUPA, Department of Physics, University of Strathclyde, }\\
\normalsize{Rottenrow East, Glasgow G4 0NG, UK.}\\
\\
\normalsize{${}^{\ast}$Corresponding author: kd1@st-andrews.ac.uk}
}

\date{}

\begin{document} 


\baselineskip24pt


\maketitle 




\begin{sciabstract}
Optical approaches to fluorescent, spectroscopic and morphological imaging have made exceptional advances in the last decade. Super-resolution imaging and wide-field multiphoton imaging are now underpinning major advances across the biomedical sciences. Whilst the advances have been startling, the key unmet challenge to date in all forms of optical imaging is to penetrate deeper. \hll{A number of schemes implement aberration correction or the use of complex photonics to address this need. In contrast, here we approach this challenge by implementing a scheme that requires no \textit{a priori} information about the media nor its properties.} Exploiting temporal focusing and single-pixel detection in our innovative scheme we obtain wide-field two-photon images through various turbid media including a scattering phantom and tissue \hll{reaching a depth of up to seven scattering mean-free-path lengths. 
Our results show that it competes favorably with standard point-scanning two-photon imaging, with up to a five-fold improvement in signal-to-background ratio while showing significantly lower photobleaching.} 
\end{sciabstract}



\section*{Introduction}
A suite of powerful, disruptive optical imaging approaches across the physical and biomedical sciences \hll{has recently emerged.} \hll{Super-resolution imaging} led to new studies looking at nanometric features within cells that have revealed intricate aspects of sub-cellular processes~\cite{Betzig2006,Hell1994,Chen2014,Pullman2016,Westmoreland2016}.
At the larger scale, methods such as optical coherence tomography~\cite{Huang1991} and light-sheet imaging~\cite{Power2017} are taking hold in fields such as opthalmology, neuroscience and developmental biology. In tandem with the requirement for a fast, wide-field visualization and super-resolved imaging across biomedicine, a grand challenge is to perform such imaging through highly scattering (turbid) media, namely tissue. In particular, this is essential to move from superficial surface imaging to functional imaging at depth~\cite{Bertolotti2012,Papagiakoumou2013,Katz2014,Kang2015}, 
which is crucial for biomedical areas including neuroscience. 
To address this area, aberration correction can be implemented~\cite{Booth2014}. However, this does not readily take into account the properties of the medium and actual retrieval of the emitted signal from depth in the medium can still be challenging. Key advances have emerged by a consideration of the propagation of light within a complex media. In this field, a number of approaches use dynamic wavefront shaping for illumination of the sample with a calculated input complex wavefront~\cite{Debarre2009,Rueckel2006,Cizmar2010, Mosk2012} which can focus light upon an embedded guide star. In essence, one determines the transmission matrix of the sample in this process~\cite{Popoff2010,Kang2015,Badon2016}. Whilst this is powerful, the requirement of a guide star restricts the approach. Furthermore, it requires determination of the properties of the medium at one or more individual points making it very challenging to implement for wide-field imaging.

\hll{An important advance} would be the realization of a fast wide-field imaging approach that would deliver and retrieve light from any given plane within a sample, even in the presence of turbidity. This would be \emph{without} the requirement to characterise or even actively correct the aberrating effect of the turbid medium. Our approach to achieve this goal exploits temporal focusing (TF) microscopy~\cite{Oron2005,Zhu2005}. 
By using the temporal rather than spatial degree of freedom, \hll{scanning of the optical axis for image reconstruction is avoided}. Consequently, TF may record wide-field multiphoton images~\cite{Oron2005,Schrodel2013}. In addition, a little recognised facet of TF is its ability to deliver light through scattering media. This has been used to project optical patterns for applications such as optogenetics, providing photo-stimulation at remarkable depths~\cite{Papagiakoumou2013,Rowlands2016,Pegard2017}. 
Although TF can deliver light through a scattering medium very efficiently, collecting the emitted fluorescent light back through the same medium - that is, truly achieving imaging  - has not been accomplished to date.
Separately, there has been the emergence of single-pixel detection, sometimes termed computational ghost imaging~\cite{Padgett2017}. In this form of imaging, known patterns illuminate an object, a single-element photodetector records the light intensity that is either transmitted or backscattered by the object and images are reconstructed with the appropriate algorithm~\cite{Tajahuerce2014,Duran2015}. 

However, whilst these studies in TF microscopy and in single-pixel detection have shown promise, none of this work has addressed the challenge of correction-free wide-field imaging through turbid media. The scheme we present here, which we call TempoRAl Focusing microscopy with single-pIXel detection (TRAFIX), uses a judicious combination of TF illumination with single-pixel imaging to obtain \hll{wide-field images of fluorescent microscopic samples within or even beyond biological tissues, in the presence of multiple scattering,} without aberration correction or characterisation of the turbid medium.

\section*{Results}
\subsection*{Principle of the technique}
\begin{figure}[htbp]
\centering
\includegraphics[scale=0.55]{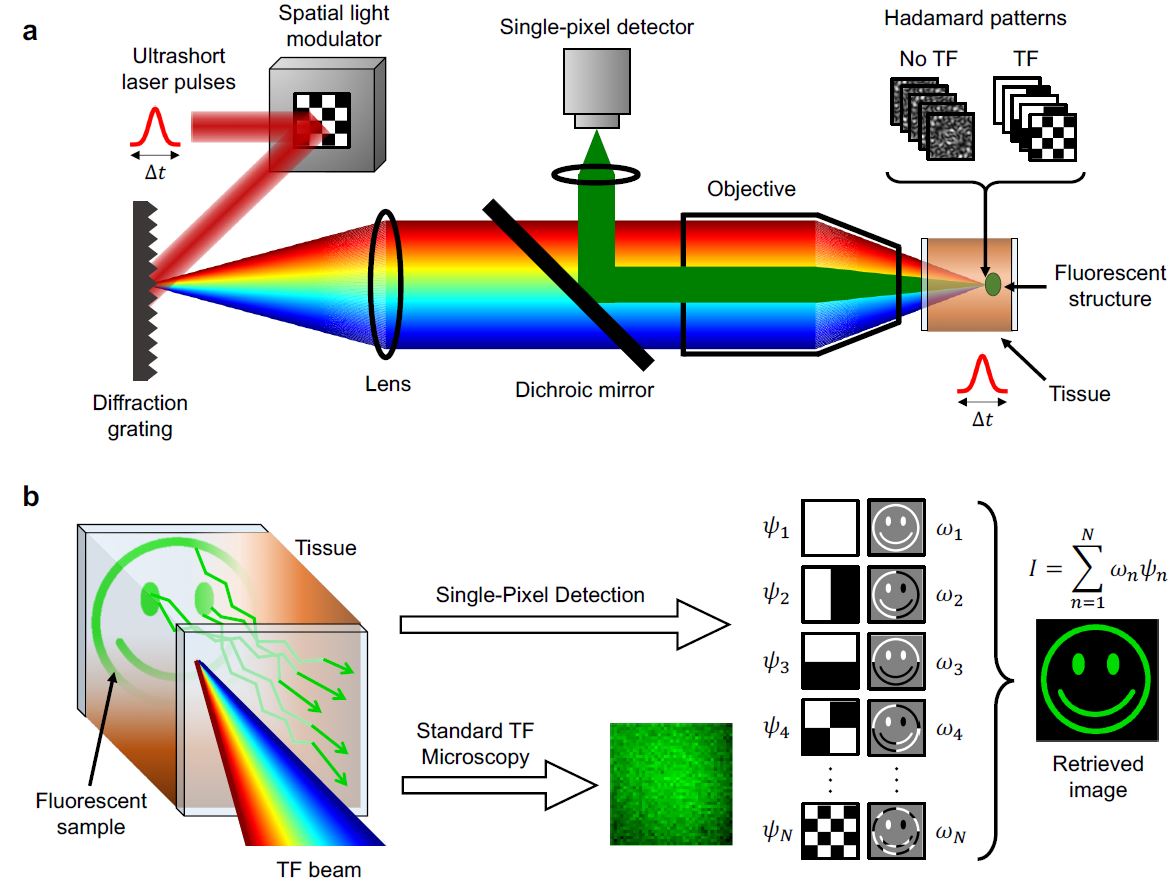}
\caption{\textbf{Working principle of TRAFIX.} (\textbf{a}) A femtosecond laser beam is expanded onto a spatial light modulator (SLM) that generates Hadamard patterns. Subsequently, the beam is diffracted from a grating and the Hadamard patterns are projected onto a fluorescent sample after propagating through a scattering medium. Fluorescent light emitted by the sample is collected by the same objective after passing through the scattering medium a second time (epi-fluorescence geometry) and the total intensity is measured by a single-pixel detector. (\textbf{b}) A TF beam propagates through a turbid medium with minimal distortion retaining the integrity of illumination patterns in the sample plane. Emitted fluorescent photons scatter as they propagate back through the tissue. In contrast to standard TF microscopy, TRAFIX tolerates scrambling of back-propagating light since only an intensity measurement is performed. 
In a single-pixel measurement, the fluorescent target is sequentially illuminated with Hadamard patterns ($\psi_{n}$) and the total intensity detected is stored as a coefficient ($\omega_{n}$). Gray background in the second column denotes regions of 0 intensity. By adding up the Hadamard patterns weighted by their respective coefficients, an image of the fluorescent sample is reconstructed.}
\label{Diagram}
\end{figure}

TF is based on decomposing an incident ultrashort pulsed light field into its constituent wavelengths with a diffraction grating. Each wavelength propagates along an individual path in the optical system and these constructively recombine to regain the original pulse duration only at the plane conjugate to the grating, generating axially confined mutiphoton excitation. 
In TRAFIX, orthonormal light patterns (in a Hadamard basis) are temporally focused through a turbid medium to illuminate a fluorescent microscopic sample of interest.
The use of TF for this projection ensures the retention of the integrity of these patterns at any given plane within the turbid media (Fig. \ref{Diagram} (a)). 
This can be regarded as due to the fact that ballistic photons remain unperturbed all the way to the object plane and arrive at the same time, contributing to the reconstitution of the pulse.
In addition, the superposition of wavelets of slightly different wavelengths at the focal plane results in nearly speckle-free propagation through long distances in scattering media as recognised by E. Papagiakoumou \textit{et al.} \cite{Papagiakoumou2013}. We confirm these aspects here with a numerical simulation. The same principle has been previously used for reducing out-of-focus excitation in \hll{line-scanning two-photon microscopy~\cite{Zhu2005}.} 
A scattering medium may effect the spatial and temporal degrees of freedom of an input field differently. In the time domain, the temporal profile of femtosecond pulses is not significantly distorted at substantial imaging depths such as 1-mm-thick brain tissue~\cite{Katz2011}. As a consequence TF may induce much more efficient multiphoton excitation when compared to standard point-focusing where spatial speckle greatly reduces the photon density at the focal spot. Consequently, TF is more robust than conventional focusing resulting in a more intense fluorescence signal generated at large depths~\cite{Sun2015} which is a major attribute for our approach.


The total intensity emitted by the fluorescent sample under each illumination pattern is collected by the same objective after passing a second time through the scattering material, in a configuration reminiscent of a single-pixel imaging. In this way we remove the requirement for any spatial resolution on the imaging path which in turn means we can readily tolerate the scrambling of the emitted fluorescence through the scattering medium (Fig. \ref{Diagram} (b)). We retain exact spatial information of where the sample is illuminated by virtue of using patterned illumination. \hll{This allows an original form of TF microscopy to be realised, enabling the use of the full penetration capabilities of TF beams for imaging at depth~\cite{Rowlands2015}.}

In the present experiments, \hll{an illumination laser with central wavelength at 800 nm delivers 140 fs pulses (80 MHz repetition rate, average output power up to 4W) onto the sample and the emitted fluorescent} photons are detected by an EMCCD camera which is used as a single-pixel detector. The epi-fluorescence configuration of TRAFIX makes it readily suitable for a suite of biomedical applications. An additional microscope takes reference images of the fluorescent sample in a transmission geometry analogous to previous reports~\cite{Rowlands2016,Papagiakoumou2013}. Reference images are taken with a CCD camera under uniform TF illumination across the field of view (FOV) (Materials and Methods). We stress that this additional reference system is not required for imaging. Once all patterns have been sequentially projected and their intensity coefficients measured, images are reconstructed using an orthogonal matching pursuit (OMP) algorithm~\cite{Tropp2007}. The OMP algorithm determines which patterns contribute most effectively to the image reconstruction and sums them up to create an image (Supplementary Materials). The number of pixels in the retrieved image is determined by the size of the Hadamard basis used in the measurement. An $n\times n$ pixels image requires a Hadamard basis containing $N = n^2$ patterns. Therefore, depending on the pixel resolution required, different number of Hadamard patterns (typically 4,096 or 1,024) are encoded on a spatial light modulator (SLM). The acquisition time of the microscope is given by $T = 2 n^{2} (t_{exp} + t_{SLM})$, where $t_{exp}$ is the exposure time of the camera used as a single-pixel detector and $t_{SLM}$ is the time required to refresh the Hadamard patterns on the SLM (including data transmission).



As TRAFIX is based on patterned illumination, it lends itself to compressive sensing measurements~\cite{Duran2015, Candes2008}. One of the main advantages of compressive sensing is that sparse signals can be reconstructed with fewer samples than required by Nyquist sampling theory. In terms of microscopy, it means that one does not need to measure with the full set of Hadamard patterns to obtain a good quality image. The compression ratio CR = N/M denotes how many patterns are used to reconstruct the image in relation to the total number of patterns in the Hadamard basis~\cite{Duran2015}. Here, M is the number of patterns used in the reconstruction algorithm. For example, a 64$\times$64 pixels image requires a measurement with a Hadamard basis containing 4,096 patterns. Consequently, a CR of 2 corresponds to using only half of the total patterns to reconstruct the image, i.e. 2,048 patterns, a CR = 4 uses only 1,024 patterns and so on. 

As we describe below, to demonstrate the performance of TRAFIX we imaged various microscopic fluorescent samples making use of full Hadamard bases to obtain a high quality image. Additional compressed images were obtained \textit{a posteriori} to demonstrate that compressive sensing measurements are possible in this configuration (Supplementary Materials). 




\subsection*{Imaging through scattering media}


To begin with, 400 nm diameter green fluorescent beads and fixed Human Embryonic Kidney (HEK293T/17) cells labeled with Green Fluorescent Protein (GFP) were imaged through scattering phantoms, designed to mimic the scattering properties of biological tissue, and through unfixed human colon tissue. A custom-made fluorescent microstructure was then imaged through fixed rat brain tissue. As scattering clearly dominates over absorption in the range of wavelengths considered in our investigation~\cite{Cheong1990,Yaroslavsky2002}, we use scattering mean free path ($l_s$) as a reference value to quantify imaging depth. The $l_s$ for the scattering media used in the experiments presented here are approximately 140~$\upmu$m, 85~$\upmu$m and 55~$\upmu$m for the scattering phantom, colon tissue and brain tissue, respectively (Materials and Methods). Full Hadamard bases of either 32$\times$32 or 64$\times$64 pixel resolution were projected onto the samples and the resulting images were reconstructed with different CR. The lateral resolution of the microscope is defined as twice 
the pixel size in the reconstructed images\hll{, and thus it depends on the FOV}. Using a Hadamard basis containing 32$\times$32 pixels \hll{and a FOV of 90$\times$90 $\mum^2$,} the resolution is 5.6 $\upmu$m and for a larger basis of 64$\times$64 pixels it is 2.8 $\upmu$m (Fig. \ref{fig:Dog}). The depth resolution in the absence of any scattering layer \hll{for a 40$\times$ NA = 0.8 objective} was measured to be 4.7~$\pm$~0.5~$\upmu$m (Fig. \ref{fig:Depth_Profiles}) and the microscope's performance in imaging beads and HEK cells was initially tested without scattering (Supplementary Materials).

In order to quantify image quality, the signal-to-background ratio (SBR) was measured for all images presented in this work (Materials and Methods). All values are summarized in Table \ref{SNR}. 
To assess the effectiveness of this technique in photolocalization, the spacing between fluorescent beads and the size of cells in the reference image and in the reconstructed images were estimated and their deviation from the reference value was then calculated (Supplementary Materials).

\begin{figure}[htbp]
\centering
\includegraphics[width=\linewidth]{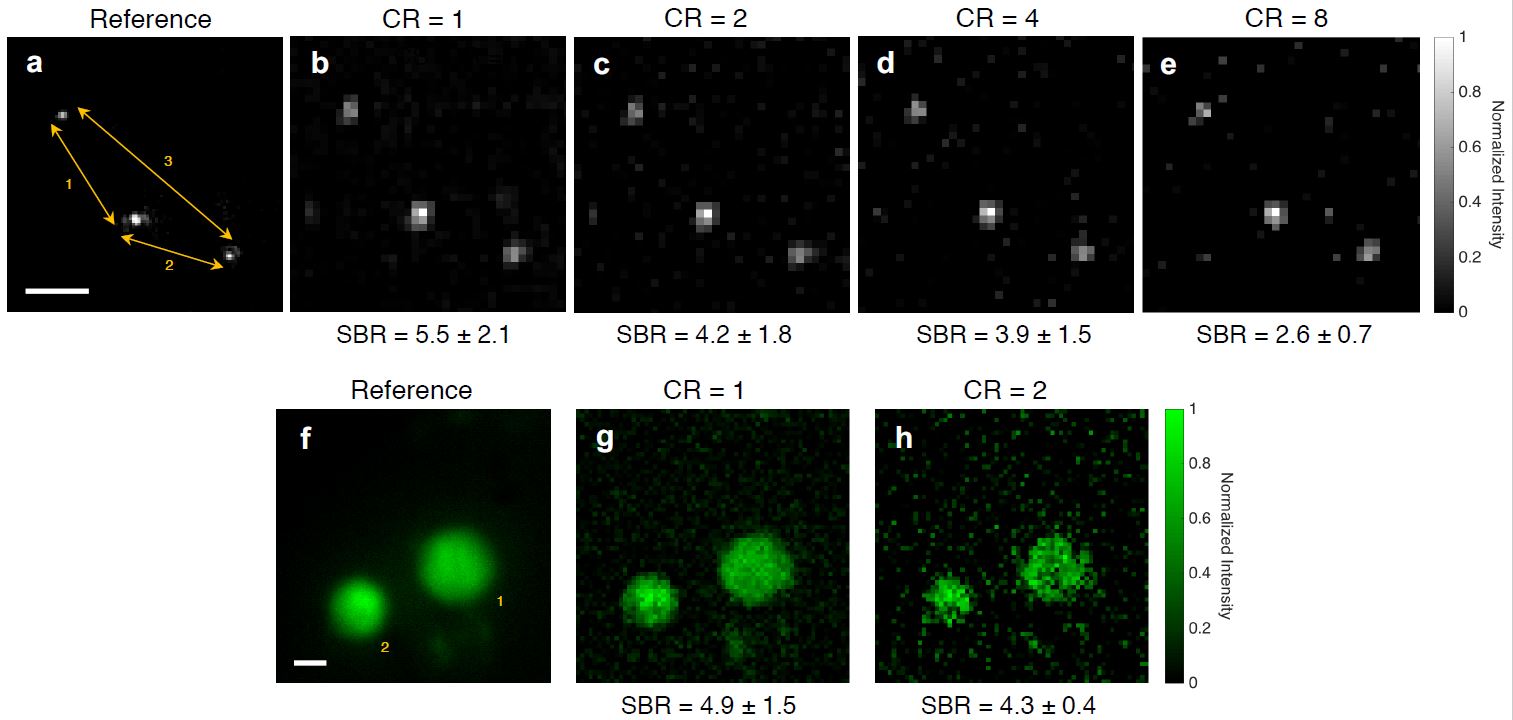}
\caption{\textbf{Images of fluorescent microscopic samples through scattering phantom.} Fluorescent beads of 400 nm in diameter and fixed HEK293T/17-GFP cells were imaged through 500 $\upmu$m and 540 $\upmu$m thicknesses of scattering phantom, respectively. (\textbf{a}, \textbf{f}) Images taken from the reference imaging system under uniform TF illumination across the FOV. Exposure time was set to 20 s and camera binning was 4 and 2, respectively. (\textbf{b}-\textbf{e}, \textbf{g}-\textbf{h}) Images obtained in epi-fluorescence configuration with TRAFIX using a Hadamard basis containing 4096 illumination patterns. They were reconstructed using different compression ratios corresponding to 100\% (CR = 1), 50\% (CR = 2), 25\% (CR = 4) or 12.5\% (CR = 8) of the total patterns. Each measurement under individual illumination patterns was taken with a binning of 64 and an exposure time of 0.5 s. The spacing between beads was measured in all five images obtaining deviations smaller than 3\% from the reference image (Table \ref{Table_Beads}). The diameter of the cells in (\textbf{f}) was measured to be 20.7 $\upmu$m and 14.3 $\upmu$m, respectively, and their values in (\textbf{g}) and (\textbf{h}) differ less than 4\% and 12\% from the reference value (Table \ref{Table_Cells}). The signal-to-background ratio (SBR) is shown for all reconstructed images. Scale bars are 10 $\upmu$m.}
\label{fig:Phantom}
\end{figure}

Figure \ref{fig:Phantom} shows images of the fluorescent beads and HEK cells obtained through 500 $\upmu$m and 540 $\upmu$m of scattering phantom (3-4 $l_s$), respectively. As a result of the strong scattering, very few illumination photons reach the sample plane generating a very dim fluorescence signal, therefore a long exposure time or large binning is required to even obtain a good reference image under uniform TF illumination in transmission. In standard TF microscopes, those few emitted fluorescence photons would need to travel back through the thick scattering medium losing all their spatial information and hence resulting in uniform noise on the camera~\cite{Rowlands2015}. Consequently, the SBR would essentially be unity, which means no signal can be extracted from the background noise. However, in TRAFIX there is no need to measure spatial information as total intensity coming from the sample under each illumination pattern is detected and used as a coefficient in the reconstruction algorithm. This results in a significant enhancement in signal detection and higher SBR values.

The achievable level of compression depends on the sparsity of the image. If an image contains very little information, high compression is possible with negligible information loss. As the images of fluorescent beads are sparse, image quality is preserved for large values of CR. A good image can be faithfully reconstructed with CR = 8 (12.5\% of the total number of patterns)
. 
Similar CR have been achieved in the literature, though with different optical imaging systems~\cite{Duran2015}. Nevertheless, we did not achieve a high compression with cells in the present configuration. They covered a larger area of the image, and thus cannot be compressed as much without a significant loss of information with the present algorithm and pixel resolution. In addition, the 800 nm excitation laser used in the experiments is not optimal for excitation of GFP, which has its highest absorption in the range of wavelengths between 870 and 920 nm~\cite{Drobizhev2011}. As a result, the cells appear very dim. 
Despite these issues, a SBR exceeding 4 was achieved at CR = 2, i.e. using only 50\% of the full Hadamard basis in the image reconstruction algorithm.

\begin{figure}[htbp]
\centering
\includegraphics[width=0.55\linewidth]{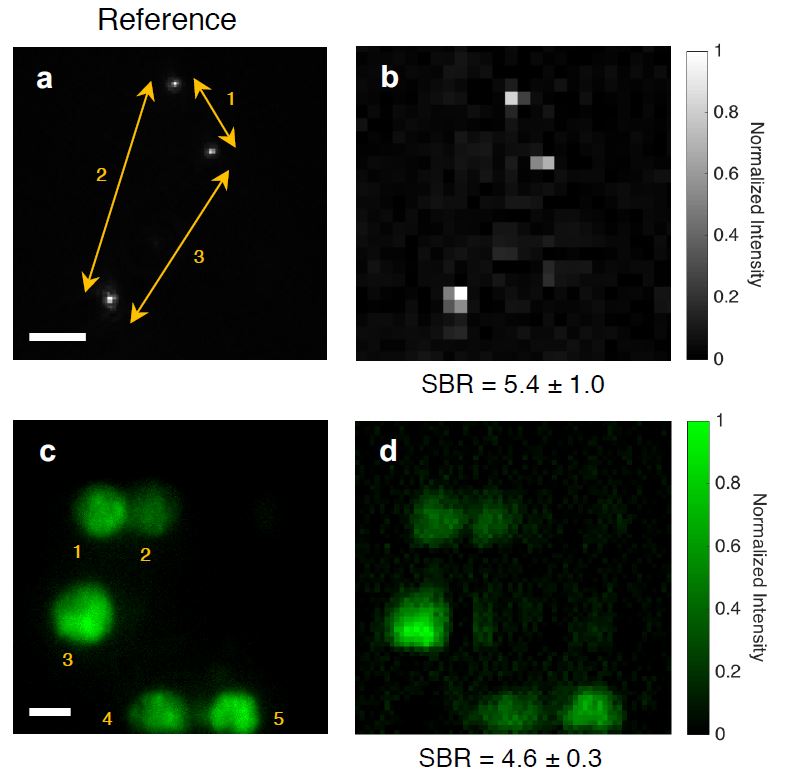}
\caption{\textbf{Images of fluorescent microscopic samples through unfixed human colon tissue.} Fluorescent beads of 400 nm in diameter and fixed HEK293T/17-GFP cells were imaged through 250 $\upmu$m and 200 $\upmu$m of human colon tissue, respectively. (\textbf{a}, \textbf{c}) Images taken from the reference imaging system under uniform TF illumination across the FOV. Camera binning in (\textbf{a}) was set to 4 and exposure time was 5 s. Camera binning in (\textbf{c)} was set to 1 and exposure time was 15 s. (\textbf{b}, \textbf{d}) Images obtained with TRAFIX using a Hadamard basis containing 1024 and 4096 illumination patterns, respectively. All patterns were used for image reconstruction (CR = 1). Camera binning for each Hadamard pattern was set to 64 and exposure time was (\textbf{b}) 1 s and (\textbf{d}) 0.75 s. The spacing between beads and the diameter of cells were measured to assess image quality (Tables \ref{Table_Beads} and \ref{Table_Cells}). The signal-to-background ratio (SBR) is shown for all reconstructed images. Scale bars are 10 $\upmu$m.}
\label{fig:Colon}
\end{figure}

In the next stage, the performance of TRAFIX was tested with unfixed human colon tissue as a scattering medium. One of the main problems in imaging through biological tissue is autofluorescence. S. Coda, {\it et al.} showed the single-photon excitation spectrum of human colon tissue at different wavelengths~\cite{Coda2014}. Under 435 nm single-photon excitation, colon tissue presents a very intense autofluorescence signal at the green part of the spectrum overlapping with the light emitted from the beads or cells used in our experiment. Single-photon excitation at 435 nm is relatively similar to two-photon excitation at 800 nm so it poses a big obstacle in the experiments because it reduces the number of photons that can reach the beads or HEK cells and it also generates undesired background light. The optical sectioning capability of TF would seem to circumvent this impediment; however, as temporally focused laser pulses propagate longer distances through scattering tissue, the resulting excitation plane becomes thicker (Fig. \ref{fig:Depth_Profiles}) \cite{Dana2011} and consequently, some autofluorescence is excited in the colon tissue resulting in high noise levels even in the reference images taken in transmission. Despite the intense autofluorescence light emitted by the colon tissue, we succeeded in imaging both 400 nm fluorescent beads and HEK cells through 250 $\upmu$m and 200 $\upmu$m ($\sim$3 $l_s$), respectively, obtaining high SBR values (Fig. \ref{fig:Colon}). \hll{An additional image taken in a scattering phantom with fluorophores extending its entire volume confirms that TRAFIX can be used in presence of out-of-focus background fluorescence (Fig. \ref{fig:3Dsample_beads}). }

The photolocalisation analysis \hll{looked at spacing between adjacent beads and cell size in a given image. It} was satisfactory in the case of images through the scattering phantom obtaining deviations smaller than 3\% for beads in any CR (Table \ref{Table_Beads}). Acceptable results were also obtained in the case of cells for the image with no compression (Table \ref{Table_Cells}). In general, larger deviations were observed for the images through colon tissue presumably due to very low photon count reaching the detector and distortions caused by inhomogeneities in the tissue.

\begin{figure}[htbp]
\centering
\includegraphics[width=0.55\linewidth]{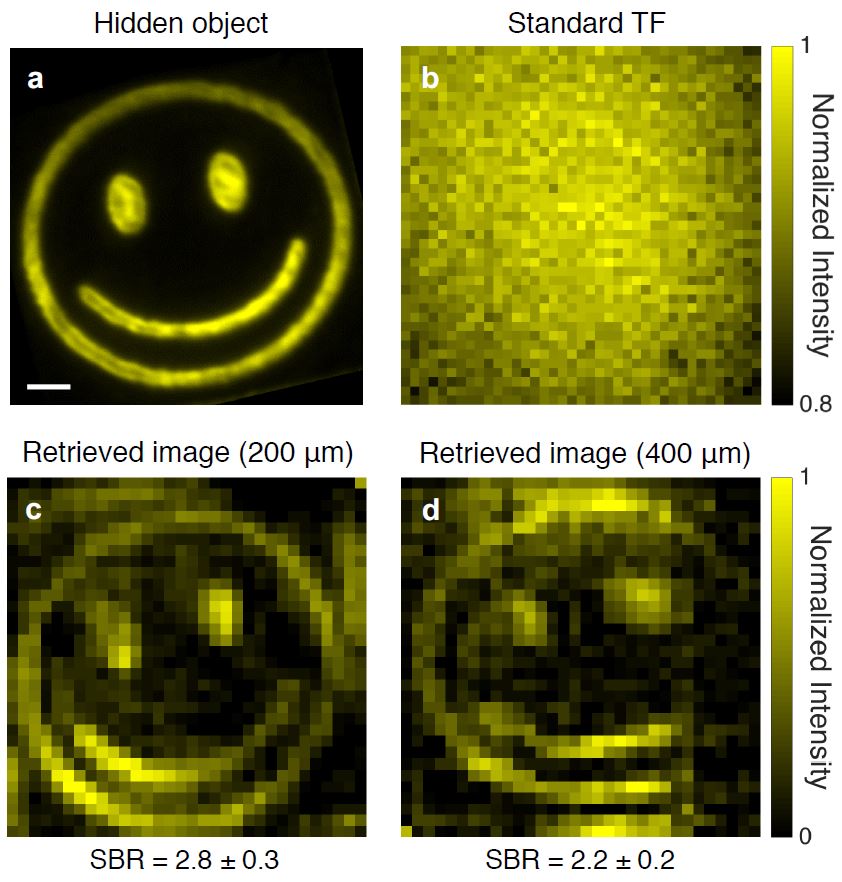} 
\caption{\textbf{Comparison of a hidden object and the retrieved images through fixed rat brain tissue.} (\textbf{a}) Reference image of a fluorescent micropattern without any scattering sample. (\textbf{b}) Image obtained using conventional TF microscopy, i.e. under uniform wide-field TF illumination with wide-field detection in epi-fluorescence configuration, through 400 $\upmu$m of fixed rat brain tissue. (\textbf{c}, \textbf{d}) Reconstructed images obtained with TRAFIX through 200 $\upmu$m and 400 $\upmu$m rat of brain tissue, respectively. The two retrieved images were reconstructed using a full Hadamard basis containing 1024 patterns. 
\hll{Camera binning was set to 64 and exposure time was (\textbf{c}) 0.2 s and (\textbf{d}) 1 s. Small intensity variations in the reconstructed images arise from inhomogeneities in the fluorescent micropattern originated in the imprinting process. Larger intensity variations are due to inhomogeneities in light transmission through the highly anisotropic scattering medium. This also applies to figures~\ref{fig:Dog}, \ref{fig:SBR_smiley_depth}, \ref{fig:SBR_smiley_colon} and \ref{fig:bleach}. }The signal-to-background ratio (SBR) is shown for all reconstructed images. Scale bar is 10 $\upmu$m.}
\label{fig:Face}
\end{figure}

Fluorescent micropatterns having a more detailed structure than beads or cells were generated on a thin fluorescent film (Materials and Methods) and imaged through a scattering phantom (Fig.~\ref{fig:Dog}\hll{ and Fig.~\ref{fig:SBR_smiley_depth}), colon tissue (Fig.~\ref{fig:SBR_smiley_colon})} and fixed rat brain tissue of different thicknesses (Fig.~\ref{fig:Face}). The maximum imaging depth achieved through rat brain tissue was 400 $\upmu$m which corresponds to approximately 7 $l_s$. The improvement of TRAFIX over conventional TF microscopy becomes evident by comparing image (b) to images (c-d) in Fig. \ref{fig:Face}.

A simple point-scanning two-photon microscope (2PM) was developed to compare the performance of TRAFIX with this widely used \hll{imaging approach (Materials and Methods). In order to provide a fair comparison, two experiments were performed in which either the lateral or axial resolution of 2PM were matched to those of TRAFIX (Supplementary Materials). The laser power per unit area was adjusted to generate equivalent fluorescence intensity on the sample for both techniques. Exposure time and camera binning were set accordingly to perform imaging in analogous conditions. 
Under such conditions, images taken without scattering show equal values of SBR in the two imaging modalities (Fig. \ref{fig:SBR_smiley_depth} (a,b)). Fig.~\ref{fig:SBR_smiley_depth} shows that SBR in 2PM degrades more rapidly than for TRAFIX when imaging at depth. This trend was confirmed by imaging through 200~$\mum$ of unfixed human colon tissue under the previously defined imaging conditions (Fig. \ref{fig:SBR_smiley_colon}). Both studies show that TRAFIX achieves between 2 and 5 times higher SBR than 2PM for the samples and depths considered in this study. 
Furthermore, an additional study shows that the wide-field nature of TRAFIX combined with patterned illumination, results in at least three times lower photobleaching of the fluorescent sample compared to 2PM, even for laser power levels favourable to the latter (Fig.~\ref{fig:bleach}, Supplementary Information).}

Figure \ref{fig:patterns} shows how the focused spot of 2PM and various TF Hadamard patterns of TRAFIX get distorted at different depths through unfixed human colon tissue. After 400$\mum$, the 2PM focused spot is not discernible and turns into a complete speckle pattern. \hll{At the same depth, low-frequency Hadamard patterns retain their shape reasonably well, though having substantial intensity inhomogeneities. These results align well with the data presented by C. J. Rowlands \textit{et al.} using TF illumination~\cite{Rowlands2016}. In such conditions, the fact that fluorescence excitation in 2PM occurs outside the expected region of focal spot suggests that TRAFIX may achieve improved imaging depths compared to 2PM, although in low resolution. It should also be noted that the axial resolution through scattering media, for the current embodiment of TRAFIX, is reduced more rapidly than that of 2PM (Fig.~\ref{fig:axial}). Axial confinement in TRAFIX could be improved by relying on line-scanning TF rather than wide-field TF illumination~\cite{Durst2008}. }




\subsection*{Numerical simulation}


In order to illustrate how TRAFIX behaves in scattering media, we performed a simplified one-dimensional simulation of the imaging process. We took into account the propagation of TF Gaussian beams through a scattering medium and the detection of fluorescent light after backward propagation through the same medium (Supplementary Materials). Our simulation shows that monochromatic light propagating through a highly scattering layer such as a 400 $\upmu$m thick brain tissue, is transformed into a spatial speckle pattern, as expected. However, one important feature of TF is that each monochromatic portion of the beam propagates through a different optical path.
As identified by E. Papagiakoumou \textit{et al.}, waves traveling with very different optical paths produce an overall smoothing of the beam while slightly different optical paths produce only an anisotropic smoothing~\cite{Papagiakoumou2013}. As a result, large features of the initial beam profile are remarkably well conserved as shown in Fig. \ref{fig:tfsim_main} (a). 
Although the main source of two-photon excitation in TRAFIX is typically generated by ballistic photons, at large depths the excitation caused by scattered photons becomes important when compared to the highly attenuated ballistic light. As the TF beams are basically speckle free, scattering of light in this case does not heavily distort the Hadamard patterns, but in fact contributes into making them brighter with the only disadvantage of generating softer edges in the features of the pattern.

In the second part of the simulation, we estimated the total fluorescent light that would reach our detector with respect to the total laser power deposited on the sample surface. It is clear to see in Fig. \ref{fig:tfsim_main} (b) that it decays dramatically as the thickness of the scattering medium increases. Thanks to single-pixel detection, in our experimental measurements we could use those very few photons to form images through up to 400 $\upmu$m of fixed rat brain tissue.




\section*{Discussion and Conclusions}

The combination of patterned TF illumination with single-pixel detection achieves remarkable imaging depths for wide-field multiphoton microscopy as it provides a way of exciting fluorescent structures deeper inside turbid media than existing imaging techniques and can efficiently collect the emitted light in an epi-fluorescence configuration. We have demonstrated the effectiveness and potential of TRAFIX by imaging fluorescent beads of 400 nm diameter and fixed Human Embryonic Kidney (HEK293T/17-GFP) cells through a layer of a scattering phantom with a thickness over 500 $\upmu$m ($\sim$~4~$l_{s}$), without any aberration correction, guide star or detector placed in/behind the turbid media. We then imaged a bright custom fluorescent microstructure through rat brain tissue of hundreds of microns in thickness reaching a maximum imaging depth of $\sim$7~$l_s$. In addition, we showed that TRAFIX works well under typical biological research conditions by imaging both fluorescent beads and HEK cells through depths over 3 $l_s$ of unfixed human colon tissue\hll{ and even in the presence of intense background fluorescence.}

The main factor that limits imaging depth of TRAFIX is the penetration depth of the TF Hadamard patterns. Propagation of TF beams through very large distances in scattering media results in distortions on the illumination patterns mainly caused by refractive index changes in the sample. Such distortions generate a basis mismatch that may result in deformities in the reconstructed images~\cite{Chi2011}.
In addition, the use of a one-dimensional diffraction grating generates horizontal shifts in the illumination Hadamard patterns that could potentially be minimized by dispersing the patterns isotropically. A future embodiment using dispersion in two dimensions with two perpendicular diffraction gratings would be an improvement. Despite these present issues, TRAFIX is capable of imaging at remarkable depths with \hll{low power per unit area }over a large FOV. Importantly, its performance may be further improved by combining it with wavefront correction making it possible to maintain spatial integrity of the illumination patterns even beyond the current limits. Furthermore, the imaging depth could be significantly improved by relying on longer wavelengths and higher order multiphoton processes such as three photon excitation~\cite{Horton2013,Rowlands2016,Toda2017} (Supplementary Materials).



The results presented here align well with previous works published in the literature such as the study by C. J. Rowlands \textit{et al.}\cite{Rowlands2015}. They compared the penetration depth of standard TF microscopy with 2PM. A careful look at their measured modulation transfer functions (MTF) shows a remarkably higher contrast under TF illumination at depths up to 100-150 $\mum$ with respect to 2PM. At larger depths, the MTF of the standard TF microscope drops dramatically due to the impossibility of retaining any spatial information in the detection system. In contrast, \hll{our TRAFIX approach utilises single-pixel detection to efficiently collect fluorescent light extending the high performance of TF to deeper regions. Our comparison with 2PM, shows that TRAFIX achieves an enhancement of up to 5 times in SBR when imaging through a scattering phantom and unfixed human colon tissue. Moreover, photobleaching of the sample is substantially reduced as a result of using wide-field TF patterned illumination rather than a focused high intensity beam (Supplementary Materials).} Image resolution and SBR can be further improved by using new approaches such as digital microscanning~\cite{Sun2016}.





\hll{In this paper we also demonstrate that compressive sensing measurements are possible in TRAFIX by showing reconstructed images \textit{a posteriori}. However, in an actual compressive sensing measurement with no \textit{a priori} knowledge of the sample, choosing the most appropriate illumination patterns is critical. Since the amount of information carried by each pattern is uneven, it is important to choose wisely the order in which they are projected to optimize image quality and acquisition speed~\cite{Sun2017}. }


\hll{In the present embodiment of TRAFIX, the acquisition speed is mainly limited by the exposure time of the EMCCD camera and the slow refresh rate of the SLM (Supplementary Materials). A typical image obtained using a full basis of 1024 patterns (32$\times$32 pixels) through a moderate thickness of scattering sample currently takes $\sim$5 min. This time is increased to up to 1 hour when imaging in high resolution (64$\times$64 pixels) with a full basis through the most challenging conditions shown in this article. To speed imaging up, the EMCCD camera would be replaced with a fast, sensitive photodetector such as a photomultiplier tube (PMT) and the imaging speed would no longer be limited by the exposure time of the detector. Besides, as TRAFIX currently uses binary Hadamard patterns, the SLM may be replaced with a significantly faster digital micromirror device (DMD, which may run at tens of kHz). This, combined with the new advances in compressive imaging, suggests that frame rates for 128$\times$128 pixel images can be increased to over $\sim$30 Hz~\cite{Higham2018,Xu2018}, enabling studies in time varying turbulence~\cite{Edgar2015}. }

Since our scheme can be easily implemented in a standard multiphoton microscope, we believe that one of its main applications will be in the field of optogenetics where it would lend itself to achieve long-term simultaneous imaging and photoactivation of neuronal networks with minimal photodamage deep inside the brain. Finally, as the polarisation state of the illumination light does not change over propagation in scattering media through the range of distances normally considered for imaging~\cite{deAguiar2017} (Supplementary Materials), TRAFIX could also be combined with polarisation-resolved imaging techniques~\cite{Chen2012}.

\hll{In summary, TRAFIX is a novel approach for deep multiphoton imaging that presents an increased SBR compared to the ubiquitous 2PM while also reducing photobleaching of the sample. In addition, the almost speckle-free propagation of TF illumination patterns suggests that TRAFIX may surpass the maximum imaging depth limit of 2PM and be very beneficial for long-term biological studies, particularly in neuroscience.}




\section*{Materials and Methods}
\subsection*{Experimental setups}\label{methods-setup}
\subsubsection*{TRAFIX}
An illumination laser (Coherent Chameleon Ultra II) delivers 140~fs pulses with 80 MHz repetition rate, up to 4W average output power at a variable central wavelength between 680~nm and 1080~nm. The central wavelength of the laser is set to 800~nm for all the experiments performed. The illumination beam is expanded four times in order to cover the active area of a phase-only spatial light modulator (Hamamatsu LCOS-SLM). The SLM is then imaged onto a blazed reflective grating (1200~g/mm) using a 4f (f~=~400~mm) telescope to create wide-field TF illumination. The first diffraction order from the SLM is transmitted through an iris in the telescope while all other orders are blocked. The beam is diffracted from the grating and all wavelengths are collimated with a f~=~400 mm lens relayed onto the back focal aperture of the illumination objective.
Two different illumination objectives are used in this work. A  water dipping objective (Nikon, 40$\times$, NA~=~0.8), which is enclosed in a custom-made chamber filled with water, generates a TF illumination plane with a size of 90 $\times$ 90 $\upmu$m$^2$. The highest average laser power per unit area used in this configuration is 64~$\pm$~5 $\upmu$W/$\upmu$m$^2$. An air objective (Nikon, 20$\times$, NA~=~0.75) is used for additional studies presented in the Supplementary Materials as accordingly specified.
Backscattered fluorescent light propagates through the turbid media and is collected by an EMCCD camera run without amplification (Andor, iXon$^{\text{EM}}$+ 885) via the same illumination objective in epi-fluorescence configuration. In order to provide reference images for this paper, forwardly emitted photons from the sample are collected by a CCD camera (Andor, Clara) in transmission via a long working distance air objective (Mitutoyo, 100$\times$, NA~=~0.7). Appropriate short pass filters are used to reject illumination laser at 800~nm and transmit fluorescence below 700~nm. In contrast to other single-pixel imaging approaches~\cite{Tajahuerce2014, Duran2015}, the EMCCD camera with 64$\times$64 binning is used as a bucket detector instead of using of a single-element detector such as a PMT or an avalanche photodiode (ADP). Using high binning helps reducing the effect of readout noise. 
All objectives, samples and cameras are attached on the body of an inverted microscope (Nikon, Eclipse Ti) accordingly.

\subsubsection*{Point-Scanning Two-Photon Microscope}
The 2PM shares the same setup as TRAFIX except the diffraction grating and a lens which are replaced with a mirror to obtain a focused beam on the focal plane of the illumination objective.\hll{ A Nikon 20$\times$ NA~=~0.75 air objective is used for all experiments. A variable iris is used to adjust the size of the focused spot. An X-Y-Z motorized stage (Nano-LP200 Mad City Labs) scans the sample in a stepwise motion} across the fixed focused beam covering the entire FOV. The same binned EMCCD camera run with no amplification (Andor, iXon$^{\text{EM}}$+ 885) collects the fluorescent light emitted by sample.

\subsection*{Fluorescent and scattering samples} \label{sampleprep}
\subsubsection*{Fluorescent Layer and Fluorescent Micropattern}
A 200 nm thin layer of super-yellow polymer spin-coated on a glass cover slip was used to characterise the profile and depth resolution of the TF beam (Supplementary Materials). It was also used to generate a fluorescent micropattern that was then imaged through scattering media. The negative of a pattern of interest was encoded on the SLM and the thin fluorescent layer was placed at the focal plane of the microscope without any scattering layer. The laser power was set to the maximum and the negative pattern was photobleached on the fluorescent film. Therefore, the only portion of the FOV that remained fluorescent was exactly the desired micropattern.

\subsubsection*{Fluorescent beads}
Green Fluorescent Polymer Microspheres (Duke Scientific, G400) with a diameter of 0.39 $\upmu$m were used to test the performance of the imaging system. A very small amount of beads was deposited on a glass coverslip and placed on top of the scattering samples to image them through the turbid media.

\subsubsection*{HEK cells}

Human Embryonic Kidney (HEK 293T/17) cells labelled with Green Fluorescence Protein (GFP) were used to demonstrate the capability of the microscope in imaging real biological samples through scattering. Human Embryonic Kidney 293T/17 cell line obtained from ATCC® was cultured in Dulbecco’s Modified Eagle’s Medium GlutaMAX™-I supplemented with 10\% Fetal Bovine Serum and 1\% Pen/Strep and was transfected using TransIT®-LT1 Transfection Reagent with the lentiviral envelope vector pSD11(VSV-G) and packaging vector pSD16 to deliver  the plasmid pLenti-GFP-Puro. Regarding the sample preparation,  the HEK 293T/17 expressing GFP were replated on WPI FluorodishTM poly-D-Lysine-coated cell culture dishes at a low density to achieve ideal imaging conditions. The day after plating, prior to imaging, the HEK293T/17-GFP cells were fixed in a PBS 4\% paraformaldehyde solution. The scattering samples were attached directly on the bottom of the dishes containing the cells.

\subsubsection*{Scattering phantom}
Polystyrene beads of 1 $\upmu$m in diameter were used as scatterers to simulate a turbid sample. The beads were purchased in a 1\% concentration solution in water (Polysciences, Microbead NIST Traceable Particle Size Standard, 1.00 $\upmu$m). The solution was thoroughly stirred in a vortex mixer and then mixed with a 1\% solution of agarose in water (preheated above melting point). Agarose and beads were mixed in the vortex mixer again and placed into sample wells of variable height. The wells consisted of a 100 $\upmu$m glass slide with multiple 90 $\upmu$m vinyl spacers stacked on top of each other. An additional cover slip was placed on top to seal the well. 
The concentration of polystyrene beads in the sample was chosen to match roughly the scattering coefficient of real biological tissue \cite{Jacques2013}. Using an on-line Mie scattering calculator~\cite{Prahl2007} 
we determined the reduced scattering coefficient of our phantom to approximately be $\upmu$$'_s \approx$ 7.5 cm$^{-1}$, corresponding to a mean free path of about $l_s = (1-g)/\upmu$$'_s \approx$ 140 $\upmu$m, where \textit{g} is the anisotropy factor ($g \sim 0.9$ for most biological tissue at the wavelength considered in this investigation). \hll{Different scattering phantoms are used in some experiments in the Supplementary Materials as appropriately specified. }

\subsubsection*{Rat brain tissue}
Similarly to our previous work~\cite{Chen2017}, rat brain tissue was obtained from adult Sprague Dawley rats\hll{ in accordance with the UKAnimals (Scientific Procedure) Act, 1986}. It was fixed and then sectioned into slices at thicknesses of 200 $\upmu$m and 400 $\upmu$m. The mean free path of the rat brain tissue was estimated by measuring the ratio of the incident laser intensity $I_0$ with the intensity of the ballistic photons $I_B$ and applying an exponential law $I_B=I_0exp(-L/l_s)$ where $L$ is the thickness of the brain tissue \cite{Badon2016}. The obtained value of $l_s =$ 55 $\pm$ 9 $\upmu$m is consistent with other measurements reported in the literature for rat \cite{Oheim2001} and mouse brains \cite{Chaigneau2011}. 

\subsubsection*{Colon tissue}

Thin fragments of unfixed normal human colon tissue were used as scattering samples. They were stored in a freezer at -80 $^{\circ}$C and mounted between two cover slips. Their thickness ranged from 200 $\upmu$m to 250 $\upmu$m and their reduced scattering coefficient was approximately 12 cm$^-$$^1$~\cite{Wei2005,Bashkatov2014}, giving a mean free path of $l_s \approx$ 85 $\upmu$m. The colon tissue sample used in this study was obtained from the Tayside Tissue Bank, Ninewells Hospital and Medical School, Dundee (Tissue request no. TR000289) with appropriate ethical permission.

\subsection*{Image quality quantification}\label{imagequality}
\subsubsection*{Signal-to-background ratio}
Signal-to-background ratio is defined here as $SBR=\mu_{sig}/\mu_{bg}$, where $\upmu_{sig}$ and $\upmu_{bg}$ are the average values of the signal and the background, respectively. In the case of cells and fluorescent micropatterns, two small regions-of-interest (ROI) were defined, one containing fluorescent structures and the other corresponding to the background. In the images of beads, the highest intensity pixels in the beads were used as signal value and the maximum intensity pixels in the rest of the image as background noise. Uncertainty is given by the standard deviation of all averaged measurements. All values of SBR presented in this work are summarized in Table~\ref{SNR}.

\subsubsection*{Cell size}
As the cells used here appear approximately spherical, their diameter was estimated by taking a measurement of their size in two orthogonal directions and averaging the obtained values. Their diameter was determined by fitting a Gaussian function to the intensity values and measuring its FWHM. Table \ref{Table_Cells} shows the diameters of the cells that appear in the different images presented in this work. Deviations in cell size between reconstructed and reference images are expressed as a percentage error.

\subsubsection*{Bead spacing}
The central pixel for each individual bead was located and the distance between beads was measured in the reference image and in all retrieved images with different compression ratios (Table \ref{Table_Beads}). Deviations in bead spacing between reconstructed and reference images are expressed as a percentage error.

\section*{Supplementary Materials}
\begin{itemize}[noitemsep]
\setlength{\itemindent}{-.4in}
\item[] note~\ref{simsuppl}. Numerical simulations
\item[] note~\ref{supp-singlepixelimaging}. Single-pixel detection and compressive sensing
\item[] note~\ref{supp-characterization}. Microscope characterisation
\item[] \hll{note~\ref{supp-TRAFIXvs2PM}. Comparison between TRAFIX and point-scanning two-photon microscopy (2PM)}
\item[] note~\ref{supp-polarization}. Polarisation state evaluation
\item[] fig.~\ref{fig:tfbeams}. Numerically simulated TF laser beam propagating through 400 $\upmu$m of brain tissue.
\item[] fig.~\ref{fig:tfsimulation}. Properties of a numerically simulated TF laser beam through brain tissue.
\item[] fig.~\ref{fig:Beam_Profile}. Effect of scattering on the beam profile with and without TF.
\item[] fig.~\ref{fig:Depth_Profiles}. Depth profile of a TF beam through a scattering phantom.
\item[] fig.~\ref{fig:FWHM_Fluor} Characterisation of a TF beam through a scattering phantom.
\item[] fig.~\ref{fig:No_Scattering}. Images of fluorescent microscopic samples without scattering.
\item[] fig.~\ref{fig:Dog}. Comparison of a hidden object and retrieved images through a scattering phantom with different resolution.
\item[] \hll{fig.~\ref{fig:3Dsample_beads}. Image of 4.8 $\mum$ fluorescent beads in a volumetric scattering phantom.} 
\item[] \hll{fig.~\ref{fig:SBR_smiley_depth}. Comparison of signal-to-background ratio (SBR) of TRAFIX and point-scanning two-photon microscopy (2PM) at depth.}
\item[] \hll{fig.~\ref{fig:SBR_smiley_colon}. Comparison of TRAFIX and point-scanning two-photon microscopy (2PM) through human colon tissue.}
\item[] \hll{fig.~\ref{fig:axial}. Axial confinement in TRAFIX and point-scanning two-photon microscopy (2PM).}
\item[] \hll{fig.~\ref{fig:bleach}. Photobleaching comparison of TRAFIX and point-scanning two-photon microscopy (2PM).}
\item[] fig.~\ref{fig:patterns}. Effect of scattering on illumination beams in point-scanning two-photon microscopy (2PM) and TRAFIX.
\item[] fig.~\ref{fig:polarization}. Effect of turbid media on light polarisation.

\item[] table~\ref{SNR}. Signal-to-background ratio (SBR) measured for all the images shown in this work.
\item[] table~\ref{Table_Cells}. Cell diameters of all images shown in this work.
\item[] table~\ref{Table_Beads}. Beads spacing corresponding to all images shown in this work.
\end{itemize}

\bibliographystyle{ScienceAdvances}

\noindent \textbf{Acknowledgements:}
We thank James Glackin and Pavlos Manousiadis from University of St Andrews for preparing the uniform thin fluorescent layers of super-yellow polymer and Jonathan Nylk, Zhengyi Yang, Graham Bruce and Federico Gasparoli for useful discussions and support. 
\noindent \textbf{Funding:}
This work is supported by the UK Engineering and Physical Sciences Research Council for funding through grants EP/P030017/1 and EP/M000869/1, and has received funding from the European Union's Horizon 2020 Programme through the project Advanced BiomEdical OPTICAL Imaging and Data Analysis (BE-OPTICAL) under grant agreement no. 675512, The Cunningham Trust and The RS MacDonald Charitable Trust. KD acknowledges the financial support of Elizabeth Killick and Susan Gurney.
\noindent \textbf{Author contributions:}
AEM and RS developed the experimental setup with guidance from KD. MM and RS developed the initial code for single-pixel microscopy which AEM adapted to the presented experiments. AEM performed experiments, recorded the data presented and performed the analysis with input and guidance from KD. MC developed the numerical model with input from KD and AEM. WAS, MA and CSH provided the cell and tissue samples. AEM, MC and KD wrote the paper with input from the other authors. KD conceived and supervised the study.
\noindent \textbf{Competing interests:} 
The work has been filed as a patent by the University of St Andrews. The authors declare no other competing interests.
\noindent \textbf{Code availability:}
The codes used to analyse the findings of this study are available on request from the corresponding author K.D.. The codes are not publicly available due to the intellectual property filed with regard to the work presented.
\noindent \textbf{Data availability:}
The data that support the findings of this study are available upon request from the corresponding author K.D.. The data are not publicly available due to the intellectual property filed with regard to the work presented.

\begin{figure}[htbp]
\centering
\includegraphics[width=0.85\linewidth]{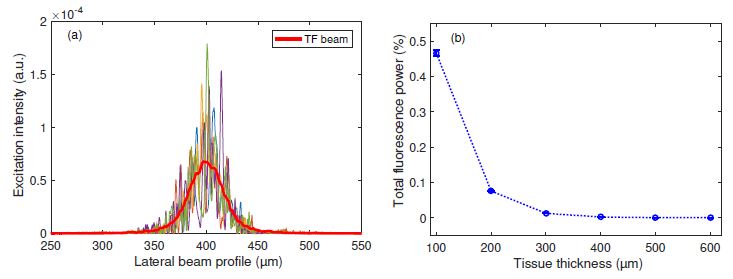}
\caption{\textbf{Numerical simulation of TRAFIX in scattering media.} \textbf{(a)} Simulated TF laser beams at the focal plane through a 400 $\upmu$m thick brain tissue. 
\hll{The solid red curve indicates the smoothed-out lateral beam profile taking all monochromatic components of the laser pulse into account.} \textbf{(b)} Total fluorescence power collected using a NA = 0.8 microscope objective for different thicknesses of brain tissue. Incident laser power at sample surface is set to 100 (a.u.).}
\label{fig:tfsim_main}
\end{figure}

\clearpage
\newpage

\begin{center}
{\large Supplementary Materials for}\\
{\LARGE Wide-Field Multiphoton Imaging Through Scattering Media Without Correction}
\end{center}

\begin{center}
\author{Adri\`{a} Escobet-Montalb\'{a}n$^1$, Roman Spesyvtsev$^{1, \dagger}$, Mingzhou Chen$^1$, Wardiya Afshar Saber$^2$, Melissa Andrews$^3$, C. Simon Herrington$^4$, Michael Mazilu$^1$ and Kishan Dholakia$^{1,\ast}$}\\
\end{center}

\setcounter{page}{1}
\setcounter{section}{0}
\setcounter{figure}{0}
\setcounter{table}{0}
\setcounter{equation}{0}

\renewcommand{\theequation}{S\arabic{equation}}
\renewcommand{\thefigure}{S\arabic{figure}}
\renewcommand{\thetable}{S\arabic{table}}
\renewcommand{\thesection}{S\arabic{section}}
\renewcommand{\thepage}{S\arabic{page}}

{\large\textbf{This PDF file includes:}}
\begin{itemize}[noitemsep]
\setlength{\itemindent}{.25in}
\item note~\ref{simsuppl}. Numerical simulations
\item note~\ref{supp-singlepixelimaging}. Single-pixel detection and compressive sensing
\item note~\ref{supp-characterization}. Microscope characterisation
\item note~\ref{supp-TRAFIXvs2PM}. Comparison between TRAFIX and point-scanning two-photon microscopy (2PM)
\item note~\ref{supp-polarization}. Polarisation state evaluation
\item fig.~\ref{fig:tfbeams}. Numerically simulated TF laser beam propagating through 400 $\upmu$m of brain tissue.
\item fig.~\ref{fig:tfsimulation}. Properties of a numerically simulated TF laser beam through brain tissue.
\item fig.~\ref{fig:Beam_Profile}. Effect of scattering on the beam profile with and without TF.
\item fig.~\ref{fig:Depth_Profiles}. Depth profile of a TF beam through a scattering phantom.
\item fig.~\ref{fig:FWHM_Fluor}. Characterisation of a TF beam through a scattering phantom.
\item fig.~\ref{fig:No_Scattering}. Images of fluorescent microscopic samples without scattering.
\item fig.~\ref{fig:Dog}. Comparison of a hidden object and retrieved images through a scattering phantom with different resolution.
\item fig.~\ref{fig:3Dsample_beads}. Image of 4.8 $\mum$ fluorescent beads in a volumetric scattering phantom. 
\item fig.~\ref{fig:SBR_smiley_depth}. Comparison of signal-to-background ratio (SBR) of TRAFIX and point-scanning two-photon microscopy (2PM) at depth.
\item fig.~\ref{fig:SBR_smiley_colon}. Comparison of TRAFIX and point-scanning two-photon microscopy (2PM) through human colon tissue.
\item fig.~\ref{fig:axial}. Axial confinement in TRAFIX and point-scanning two-photon microscopy (2PM).
\item fig.~\ref{fig:bleach}. Photobleaching comparison of TRAFIX and point-scanning two-photon microscopy (2PM).
\item fig.~\ref{fig:patterns}. Effect of scattering on illumination beams in point-scanning two-photon microscopy (2PM) and TRAFIX.
\item fig.~\ref{fig:polarization}. Effect of turbid media on light polarisation.

\item table~\ref{SNR}. Signal-to-background ratio (SBR) measured for all the images shown in this work.
\item table~\ref{Table_Cells}. Cell diameters of all images shown in this work.
\item table~\ref{Table_Beads}. Beads spacing corresponding to all images shown in this work.
\end{itemize}

\clearpage
\section{Numerical simulations}\label{simsuppl}
\subsection*{Two-photon temporal focusing microscopy} 
\begin{figure}[htbp]
\centering
\includegraphics[width=\linewidth]{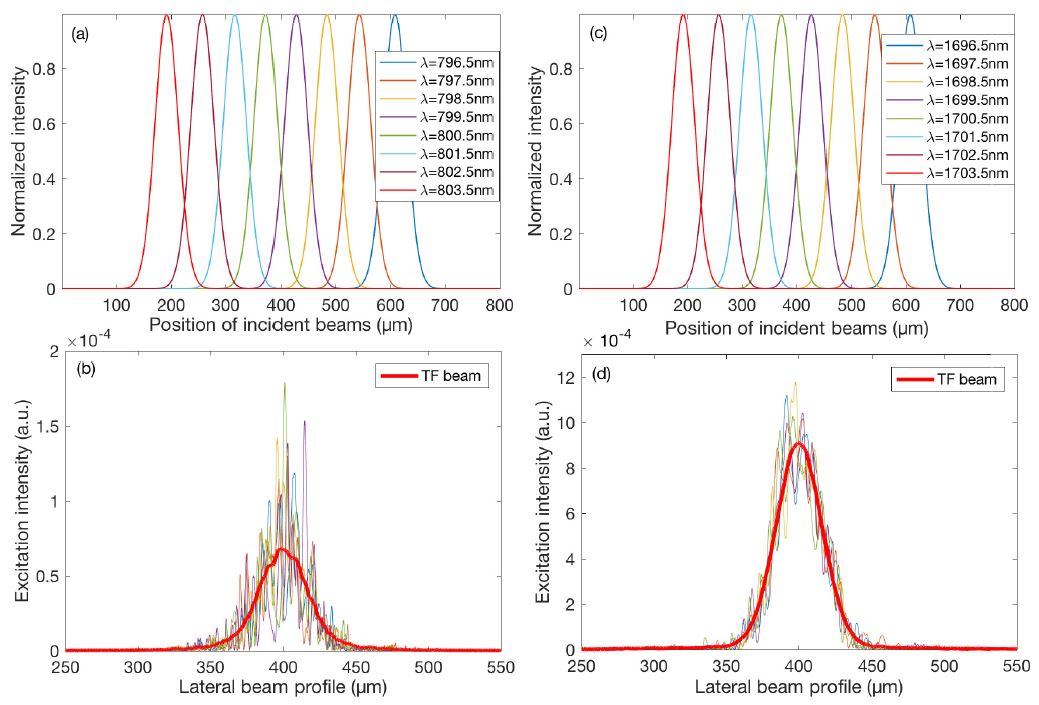}
\caption{\textbf{Numerically simulated TF laser beam propagating through 400 $\upmu$m of brain tissue.} (\textbf{a}) The laser pulse, having central wavelength at 800 nm and (\textbf{c}) central wavelength at 1700 nm, is dispersed into its constituent frequencies, each represented as a Gaussian beam (FWHM = 35 $\upmu$m), at the tissue surface. (\textbf{b}, \textbf{d}) Beam profiles at the focal plane for two- and three-photon excitation, respectively. The solid red curve represents the overall lateral beam profile after propagating through brain tissue showing how speckle pattern is considerably smoothed out.}
\label{fig:tfbeams}
\end{figure}

\begin{figure}[htbp]
\centering
\includegraphics[width=0.6\linewidth]{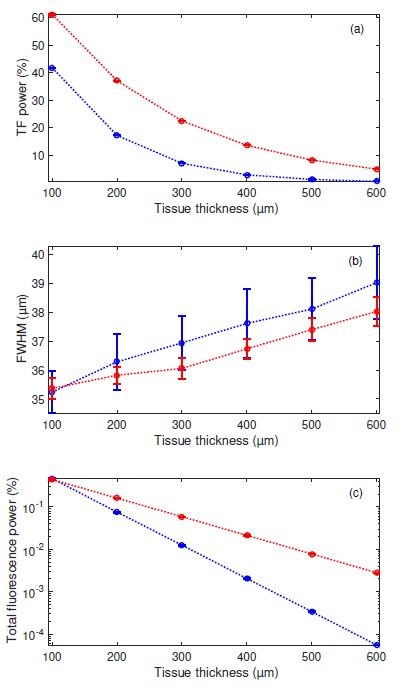}
\caption{\textbf{Properties of a numerically simulated TF laser beam} through brain tissue. \textbf{(a)} Total TF laser power at focal plane, \textbf{(b)} FWHM of a TF Gaussian beam and \textbf{(c)} total collected fluorescence power using a NA = 0.8 objective, for different thicknesses of brain tissue. Incident laser power is set to 100 (a.u.). Blue  and red curves correspond to simulated two- and three-photon fluorescence excitation, respectively.}
\label{fig:tfsimulation}
\end{figure}

We numerically simulate a TF Gaussian beam propagating through turbid media and exciting a uniform fluorescent layer. Laser pulses centered at 800 nm ($\Delta\lambda=\pm3.5$ nm) are used as an excitation source. A layer of turbid tissue is simulated by mixing an aqueous medium with randomly distributed dielectric spheres of 2 $\upmu$m at a concentration of 1 sphere per 1000 $\upmu$m$^3$. The refractive index of the scattering spheres is 0.1 higher than the surrounding media. Numerical simulations are carried out only in one dimension to reduce the required computation power. Similarly to the grating in the experiment, the spectral components in the laser pulse are dispersed and expanded to  fill the back aperture of a microscope objective (NA = 0.8) as shown in Fig.~\ref{fig:tfbeams} (a). Assuming that the laser pulse contains 70 spectral components at different wavelengths with a step of 0.1 nm, each monochromatic component of the pulse propagates through a different region of scattering tissue and arrives at the same position on the focal plane using the split step approach implemented in Matlab (2016a)~\cite{Chen2011,Chen2016}. Back scattering is neglected here while the attenuation coefficient of turbid media is set to $4.45\times 10^{3}$ m$^{-1}$, experimentally estimated by measuring the laser power before and after tissue samples~\cite{Flock1987}. At the focal plane, each spectral component forms a speckle pattern as shown in Fig.~\ref{fig:tfbeams} (b). These speckle patterns are 
significantly spatially correlated but shifted from each other since each different wavelength travels a slightly different optical path through the turbid medium.
As expected, temporally focusing these speckle patterns results in a distinct 'smooth' excitation beam at the focal plane as shown by the solid red curve in Fig.~\ref{fig:tfbeams} (b). 
According to the numerical simulation results, the total power delivered through the tissue decreases exponentially with the thickness of tissue as shown in Fig.~\ref{fig:tfsimulation} (a). Fig.~\ref{fig:tfsimulation} (b) also shows the beam's full width at half maximum (FWHM) at the TF plane increases slightly when the beam is focused through a thick layer which would result in a slight distortion of the Hadamard patterns projected in the experimental measurements. The excited fluorescent photons have to propagate through the turbid tissue before they can be collected by the same microscope objective and therefore the total fluorescence signal becomes very weak for a thick tissue layer compared to the initial illumination power as shown in Fig.~\ref{fig:tfsimulation} (c).  

\subsection*{Three-photon temporal focusing microscopy}
We also numerically simulated three-photon TF microscopy using laser pulses centered at 1700 nm ($\Delta\lambda=\pm3.5$ nm) to excite a fluorescent layer under the same slice of tissue. As the attenuation length at 1700 nm is almost doubled for the same brain tissue~\cite{Horton2013}, the attenuation coefficient is now set to $2.5\times 10^{3}$ m$^{-1}$. Assuming all other parameters are the same, three-photon TF beams show a better penetration through the layer of tissue. Fig.~\ref{fig:tfbeams} (d) shows that the overall smoothed speckle pattern reaching the focal plane resembles more the initial Gaussian beams than the one in (b). Additionally, thanks to using longer wavelenghts, the amount of TF laser power reaching focus is considerably higher than in the previous case Fig.~\ref{fig:tfsimulation} (a). As the FWHM of the overall smoothed beam is thinner (Fig.~\ref{fig:tfsimulation} (b)), the shape of the illumination Hadamard patterns would be distorted by a smaller amount than in two-photon excitation. This, combined to the fact that its intensity is higher (Fig.~\ref{fig:tfbeams} (d)), would make three-photon TRAFIX capable of penetrating much deeper in biological tissue. For the same fluorescence signal level detection threshold, TRAFIX in three-photon mode can penetrate nearly 1.5 times deeper into the tissue than in two-photon mode.

\section{Single-pixel detection and compressive sensing}\label{supp-singlepixelimaging}

Compressive sensing is a signal processing technique for acquiring and reconstructing signals with fewer measurements than required by the Nyquist sampling theorem. The description of single-pixel detection and compressive sensing presented in this section is based on the work done by Tajahuerce, \textit{et al.}, Dur\'{a}n, \textit{et al.}, Cand\`{e}s, \textit{et al.}, Tropp, \textit{et al.}, and Olivas, \textit{et al.} \cite{Candes2008,Olivas2013,Duran2015,Tajahuerce2014,Tropp2007}.

Any signal or image, $x$, can be expanded as a sum of weighted basis functions:

\begin{equation}\label{eq_1}
x = \sum_{n=1}^{N} \omega(n) \psi_n
\end{equation}

where $\psi_n$ are the basis functions, $\omega(n)$ are the weighting coefficients and \textit{N} is the total number of pixels. In order to represent an image with perfect accuracy, the number of basis functions used in the reconstruction should be the same as the total number of pixels in the image. It is also well known that the Nyquist sampling theorem states that the sampling rate must be at least twice the highest frequency that needs to be resolved. However, if the image is sparse, compressive sensing can break this limit and reconstruct images or signals with far fewer samples or measurements than required by Nyquist theory. Sparse images can be defined as images in which most weighting coefficients are small and only a few of them are relatively large so image quality is not lost by neglecting the smallest ones. Therefore, there is no need to measure all the basis coefficients, instead, one can make only a few random measurements that have equal probability of obtaining relevant data and with the information captured, the image can be reconstructed. In terms of Eq.~\ref{eq_1}, a good image can be reconstructed by only summing \textit{K} weighted basis functions with \textit{K} much smaller than \textit{N}:

\begin{equation}\label{eq_2}
x \cong \sum_{l=1}^{K} \omega(l) \psi_l
\end{equation}

The reconstruction process can be further optimized by using algorithms as described below.

\subsection*{Single-pixel detection and compressive sensing in TRAFIX} \label{supp-compressivesensing}

In TRAFIX, the sample is illuminated by microstructured light patterns, each corresponding to a basis function, and the total intensity emitted by the sample for each pattern is measured with a photodetector. Different sets of bases can be used to perform a single-pixel measurement. Olivas S.J., {\it et al.} compared the performance of a digital (Hadamard), a grayscale (discrete cosine transform) and a random (Noiselet) set of bases. According to their work, a random set of bases like the Noiselet, provides a more uniform sampling and higher spatial frequency information is present in the reconstructed images.  However, more images are needed to converge to a visually appealing solution. They also stated that spatially structured patterns such as the Hadamard or discrete cosine transform, interact with the structure of the sample and certain weighting coefficients stand out, which in most occasions, helps reaching more easily a solution that looks nicer. In their study, they concluded that in general, the Hadamard basis provides better results~\cite{Olivas2013}.

In our TRAFIX microscope, we performed single-pixel measurements with an \textit{a posteriori} approach to compressive sensing. We considered a basis formed of Hadamard patterns whose entries are either 1 or -1. Two important properties of Hadamard matrices are that their rows are orthogonal and they fulfill the following condition:

\begin{equation}\label{eq_3}
HH^T = nI_n
\end{equation}

where \textit{H} is a Hadamard matrix and \textit{I} an identity matrix of dimension n. Hadamard matrices were encoded on a SLM and the corresponding patterns were projected onto the sample after propagating through turbid samples. As the number of pixels in the reconstructed image is determined by the size of the basis, two different sets of Hadamard bases containing 1,024 and 4,096 elements were used to obtain images of 32$\times$32 pixels and 64$\times$64 pixels, respectively. Since Hadamard matrices are made of -1 and 1, each pattern had to be split into a sequence of two complementary patterns, one containing the positive portion and the other one the negative portion of the pattern. Then, measurements taken with the negative part of the basis were subtracted from the positive portion obtaining a weight coefficient for each element of the Hadamard basis. The background was subtracted from each measurement. Once all patterns had been projected and their weights had been determined, the final image could be reconstructed by summing them up as described in Eq.~\ref{eq_1}. However, if only a few patterns were used in the reconstruction, we could obtain different compression ratios (CR)~\cite{Duran2015}. Reconstruction with fewer patterns is \textit{equivalent} to taking very few measurements in the first place. In order to optimize the reconstruction process we implemented an orthogonal matching pursuit (OMP) algorithm in the same way as J. A. Tropp, \textit{et al.} and J. M. Phillips~\cite{Tropp2007,Phillips2017}.

\section{Microscope characterisation} \label{supp-characterization}
\subsection*{Beam profile}
\begin{figure}[htbp]
\centering
\includegraphics[width=\linewidth]{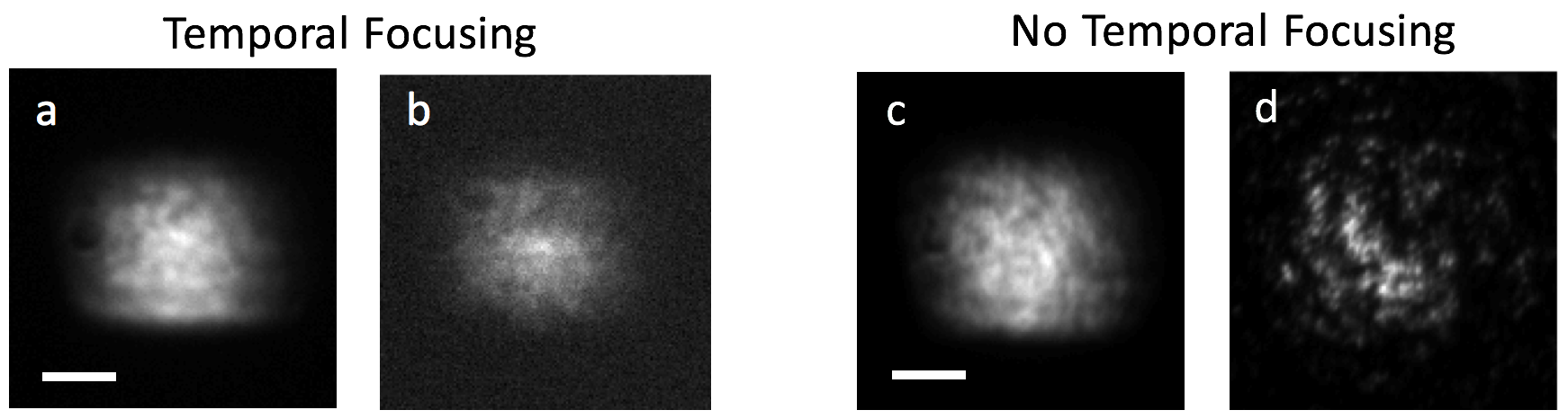}
\caption{\textbf{Effect of scattering on the beam profile with and without TF.}  (\textbf{a}, \textbf{b}) Beam profile with and (\textbf{c}, \textbf{d}) without temporal focusing. (\textbf{a},  \textbf{c}) were taken without scattering and (\textbf{b}, \textbf{d}) through 900 $\upmu$m of scattering phantom ($l_s\approx 250 \mum$). Scale bar is 20 $\upmu$m.}
\label{fig:Beam_Profile}
\end{figure}

\begin{figure}[htbp]
\centering
\includegraphics[scale=0.325]{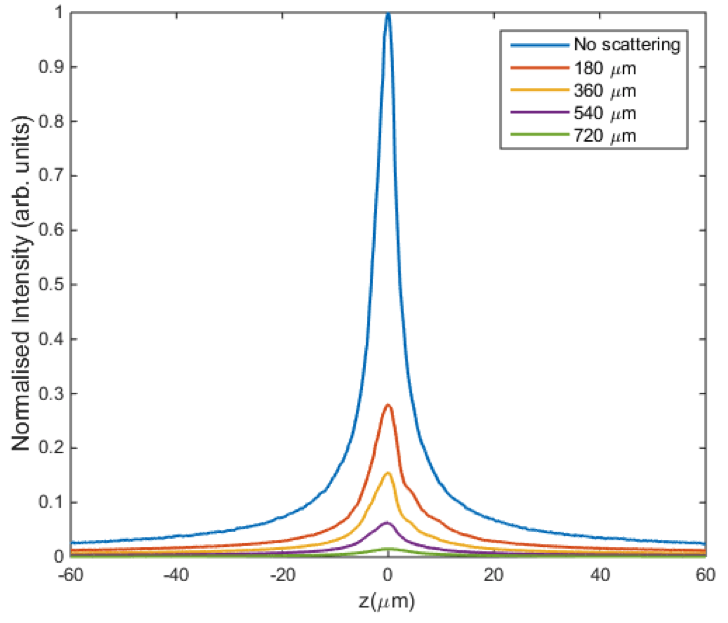}
\caption{\textbf{Depth profile of a TF beam through a scattering phantom.} The axial extent of the illumination plane was measured through different thicknesses of scattering phantom ($l_s\approx 250 \mum$). The size at the FWHM determines the depth resolution of the microscope. The axial resolution without scattering is 4.7 $\pm$ 0.5 $\upmu$m. Objective used for this measurement: Nikon 40$\times$ NA = 0.8.
}
\label{fig:Depth_Profiles}
\end{figure}

Before taking any images through scattering media, the performance of the microscope was characterised. Firstly, the profile of a TF beam was imaged on the reference arm using the fluorescent polymer layer with no scattering (Fig. \ref{fig:Beam_Profile} (a)) and with a 900 $\upmu$m thick scattering phantom ($l_s\approx 250 \mum$) (Fig. \ref{fig:Beam_Profile} (b)). 
The profile after scattering appears dimmer but its shape remains practically unchanged. Then, the diffraction grating was replaced with a mirror in order to obtain a wide-field image without TF. In this case the beam profile after scattering turns into a speckle pattern (Fig. \ref{fig:Beam_Profile} (d)). These results are in accordance with the ones presented in~\cite{Papagiakoumou2013} and show the resilience of TF to speckle formation after propagation through scattering media.

\subsection*{Axial resolution}
The axial resolution of the microscope was characterised by moving the sample axially across the focal plane with a stepper motor (Newport Universal Motion Controller/Driver, ESP300) and measuring the total intensity coming from the thin fluorescent layer at different axial positions separated 250 nm from each other. A water dipping Nikon 40$\times$ NA = 0.8 was used for this investigation. Without scattering, the thinnest excitation plane achieved was measured to be 4.7 $\pm$ 0.5 $\upmu$m. The same measurement was performed through various thicknesses of the scattering phantom, used in the previous section, to determine how the depth resolution changes as we image deeper into turbid samples (Fig. \ref{fig:Depth_Profiles}).

\begin{figure}[htbp]
\centering
\includegraphics[width=\linewidth]{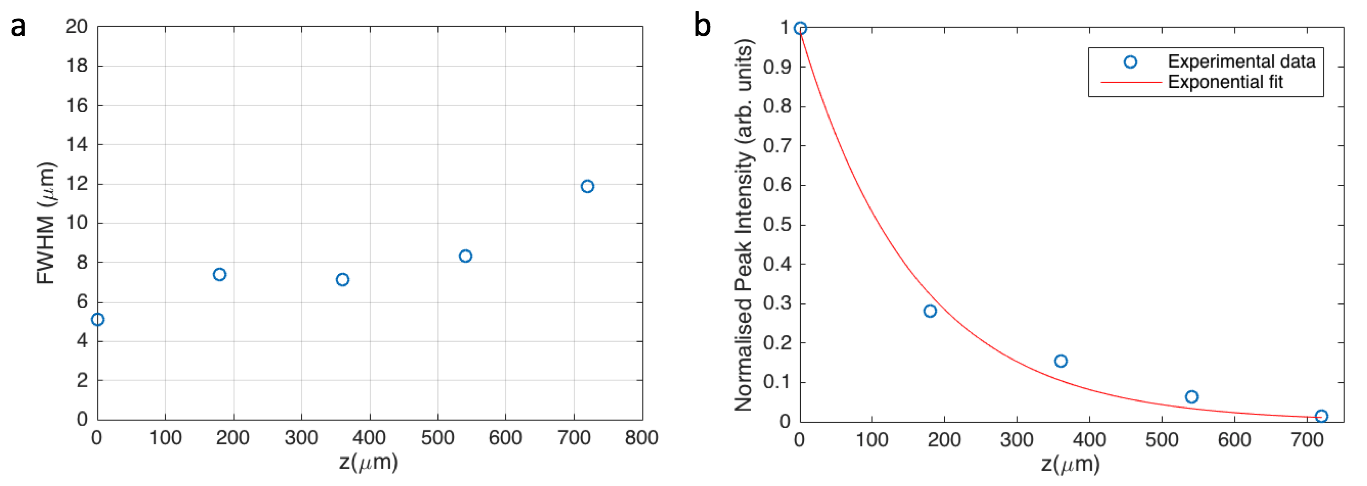}
\caption{\textbf{Characterisation of a TF beam through a scattering phantom.} Plots obtained from the data represented in Fig.~\ref{fig:Depth_Profiles}. (\textbf{a}) Full width at half maximum (FWHM) of the depth profiles. The depth resolution decreases about 2.5 times after propagating through 720 $\upmu$m of the scattering phantom ($l_s\approx 250 \mum$). (\textbf{b}) Peak intensity of the depth profiles. It decreases exponentially as light travels farther through the scattering medium.
}
\label{fig:FWHM_Fluor}
\end{figure}

\begin{figure}[htbp]
\centering
\includegraphics[width=\linewidth]{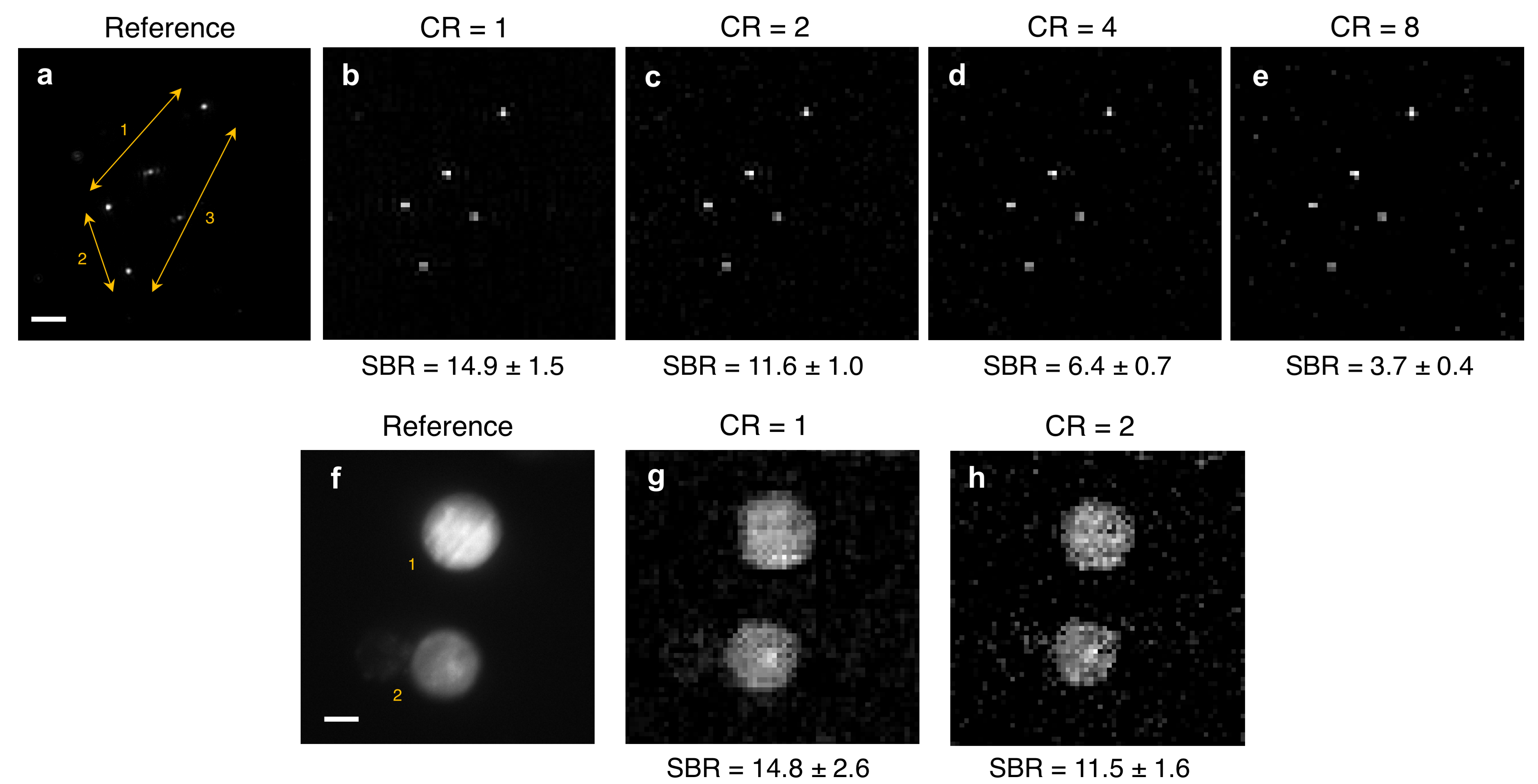}
\caption{\textbf{Images of fluorescent microscopic samples without scattering.} Fluorescent beads of 400 nm in diameter and fixed HEK293T/17-GFP cells were imaged without the presence of any scattering layer. (\textbf{a}, \textbf{f}) Images taken from the reference imaging arm under uniform TF illumination across the FOV. Camera binning was set to 1, and exposure time was 5 s and 10 s, respectively. (\textbf{b}-\textbf{e}, \textbf{g}-\textbf{h}) Images obtained in epi-fluorescence configuration with TRAFIX using a Hadamard basis containing 4096 illumination patterns. They were reconstructed using different compression ratios corresponding to 100\% (CR = 1), 50\% (CR = 2), 25\% (CR = 4) or 12.5\% (CR = 8) of the total patterns. Each single measurement of the Hadamard scan was taken with a binning of 64 and an exposure time of 0.025 s and 0.02 s, respectively. The spacing between beads (Table \ref{Table_Beads}) and the diameter of the cells (Table \ref{Table_Cells}) were measured to assess image quality. The signal-to-background ratio (SBR) is shown for all reconstructed images. Scale bars are 10 $\upmu$m.
}
\label{fig:No_Scattering}
\end{figure}

\begin{figure}[htbp]
\centering
\includegraphics[scale=0.3]{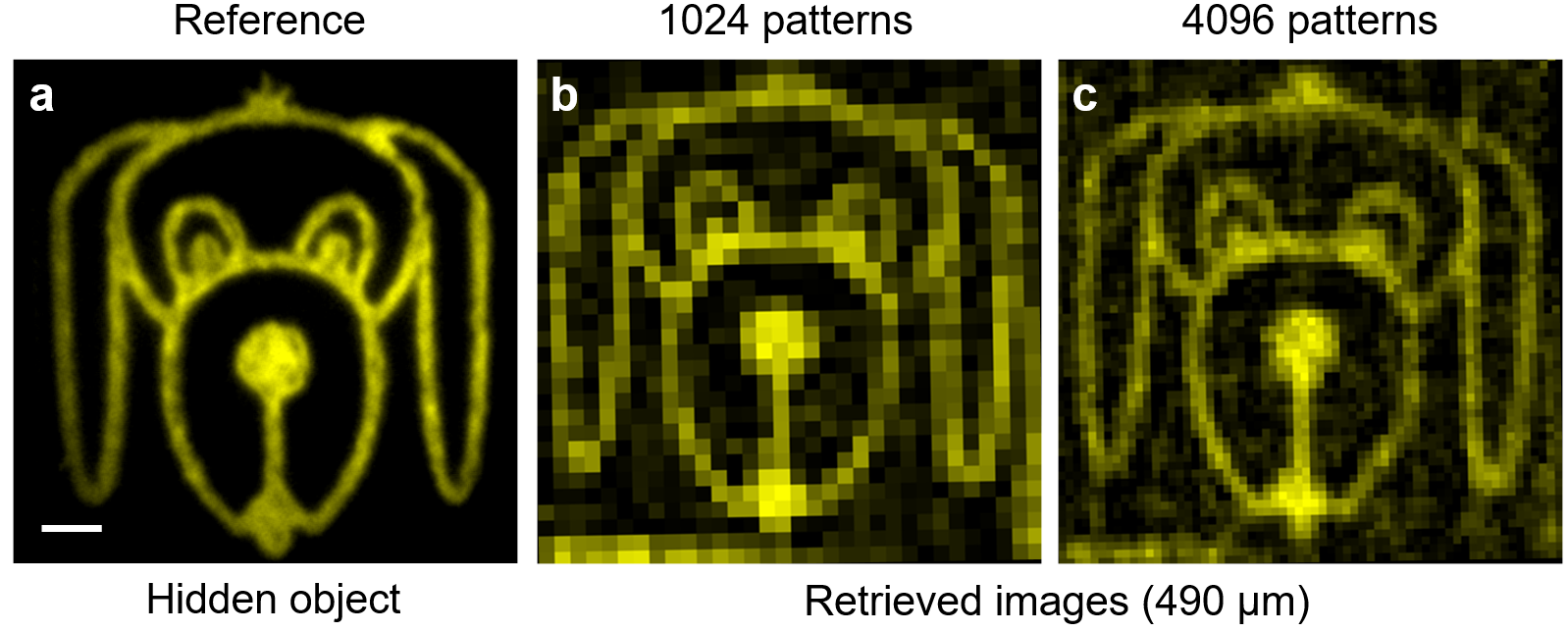}
 \caption{\textbf{Comparison of a hidden object and retrieved images through a scattering phantom with different resolution.} (\textbf{a}) Reference image of the fluorescent micropattern without any scattering sample. (\textbf{b}, \textbf{c}) Reconstructed images obtained with TRAFIX in epi-fluorescence configuration through 490 $\upmu$m of scattering phantom. The two retrieved images were reconstructed using full Hadamard bases containing 1024 and 4096 patterns, respectively. Scale bar is 10 $\upmu$m.}
\label{fig:Dog}
\end{figure}

\begin{figure}[htbp]
\centering
\includegraphics[width=0.6\linewidth]{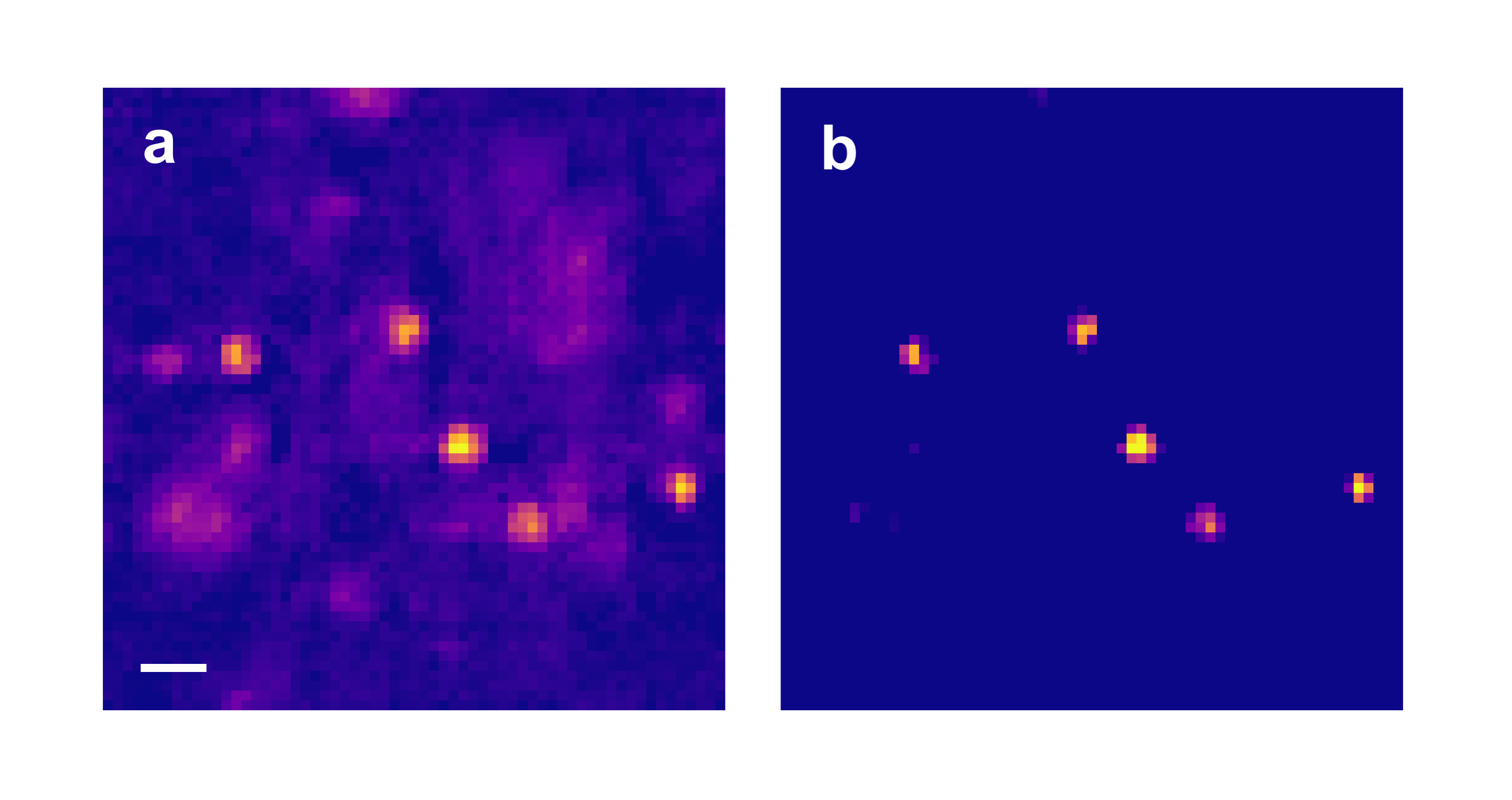}
\caption{\textbf{Image of 4.8 \boldmath{$\mum$} fluorescent beads in a volumetric scattering phantom.} 0.4 $\mum$ fluorescent beads were embedded in the phantom to generate additional background fluorescent light. (a) Image taken at a depth of $\sim 300 \mum$ with balanced gray levels. (b) Same image with gray levels adjusted to minimise background noise. Scale bar is 10 $\mum$}
\label{fig:3Dsample_beads}
\end{figure}

\begin{figure}[htbp]
\centering
\includegraphics[width=0.7\linewidth]{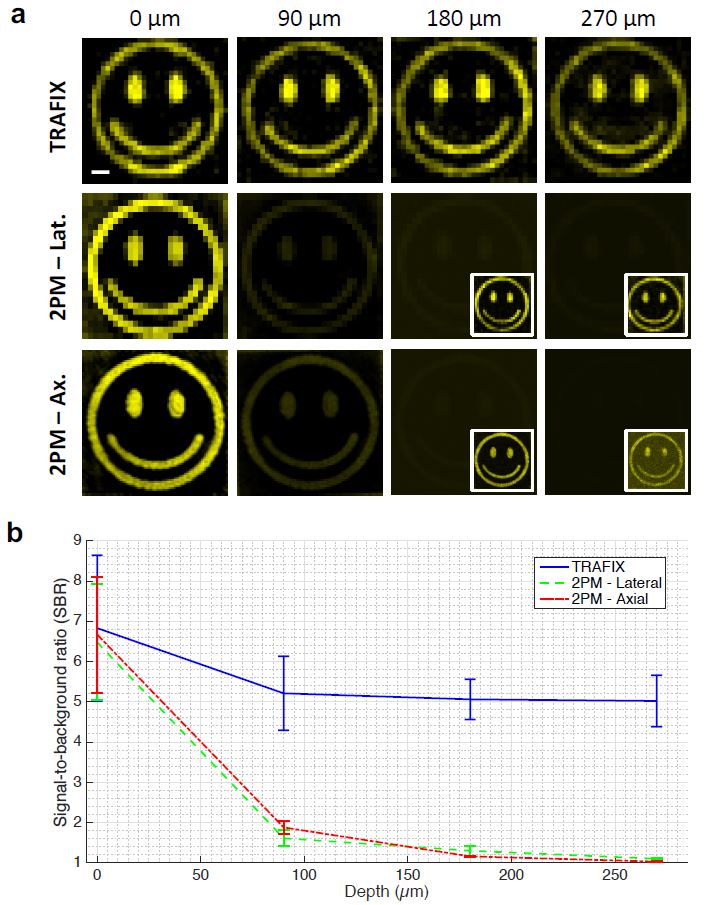}
\caption{\textbf{Comparison of signal-to-background ratio (SBR) of TRAFIX and point-scanning two-photon microscopy (2PM) at depth.} \textbf{(a)} Images of fluorescent micropatterns obtained with TRAFIX and 2PM through scattering phantoms of different thicknesses. 2PM-Lateral has approximately the same lateral resolution as TRAFIX and 2PM-Axial has similar axial confinement as TRAFIX. Small insets show the same image with different gray values. Laser power per unit area per TRAFIX pattern was 0.068 mW$/{\mum^2\cdot}$pattern and the total accumulated laser power per unit area after a full measurement with 1024 patterns was 69 mW$/\mum^2$. Laser power per unit area for 2PM was 2.2 mW$/\mum^2$. Camera binning and exposure time were maintained constant for all measurements at the same depth. Exposure time per TRAFIX pattern was the same as per pixel in 2PM. Images were acquired with an underfilled 20$\times$ NA = 0.75 objective. Scale bar is 10$\mum$. \textbf{(b)} Values of SBR at different depths for TRAFIX and 2PM. The scattering mean free path of the phantom is $l_{s}\approx 100\mum$.}
\label{fig:SBR_smiley_depth}
\end{figure}

\begin{figure}[htbp]
\centering
\includegraphics[width=0.8\linewidth]{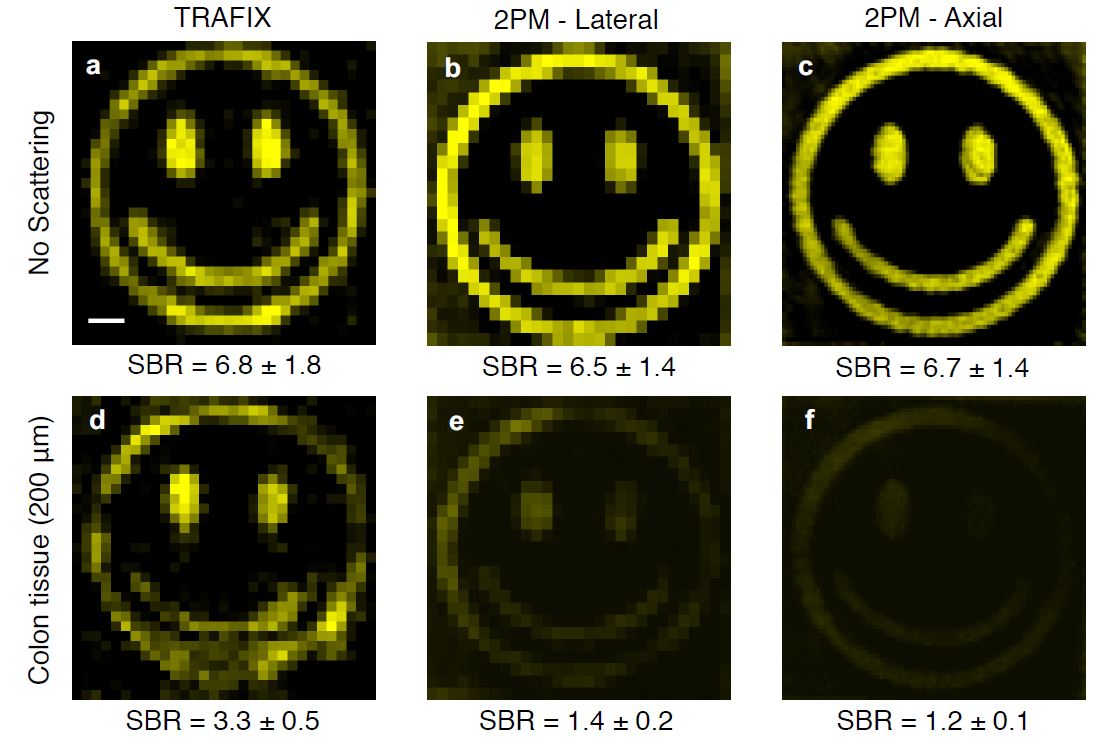}
\caption{\textbf{Comparison of TRAFIX and point-scanning two-photon microscopy (2PM) through human colon tissue.} \textbf{(a-c)} Images of a fluorescent micropattern obtained with TRAFIX and 2PM without scattering. 2PM-Lateral has approximately the same lateral resolution as TRAFIX and 2PM-Axial has similar axial confinement as TRAFIX. \textbf{(d-f)} Images obtained through $200\mum$ of unfixed human colon tissue. Laser power per unit area per TRAFIX pattern was 0.068 mW$/{\mum^2\cdot}$pattern and the total accumulated laser power per unit area after a full measurement with 1024 patterns was 69 mW$/\mum^2$. Laser power per unit area for 2PM was 2.2 mW$/\mum^2$. Camera binning and exposure time were maintained constant for \textbf{(a-c)} and \textbf{(d-f)}. Exposure time per TRAFIX pattern was the same as per pixel in 2PM. Signal-to-background ratio (SBR) is shown for all images. Images were taken with an underfilled 20$\times$ NA = 0.75 objective. Scale bar is 10$\mum$.}
\label{fig:SBR_smiley_colon}
\end{figure}

\begin{figure}[htbp]
\centering
\includegraphics[width=0.5\linewidth]{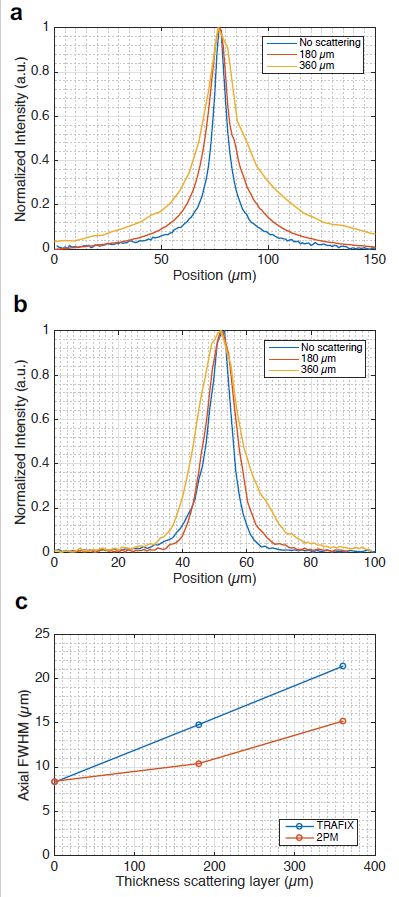}
\caption{\textbf{Axial confinement in TRAFIX and point-scanning two-photon microscopy (2PM).} Axial profile of \textbf{(a)} a uniform TF $84\times84$ $\mum^2$ square and \textbf{(b)} a spatially focused 1.26 $\mum$ diameter spot through various layers of scattering phantom. $l_s\approx 150 \mum$ \textbf{(c)} Full width at half maximum (FWHM) of the depth profile for TRAFIX and 2PM through scattering phantoms. Objective used in this experiment: Nikon 20$\times$ NA = 0.75.}
\label{fig:axial}
\end{figure}

Figure \ref{fig:FWHM_Fluor} (a) shows that the depth resolution of TRAFIX deteriorates as the beam propagates deeper inside a turbid medium but it only decreases by a factor of 2.5 after propagating through more than 700 $\upmu$m of the medium. In addition, the intensity of the light emitted by the excited plane drops exponentially (Fig. \ref{fig:FWHM_Fluor} (b)). 
The results presented in this section are in accordance with our simulations (Fig. \ref{fig:tfsim_main}) and the data presented by studies of H. Dana, \textit{et al.}~\cite{Dana2011} and E. Papagiakoumou, \textit{et al.}~\cite{Papagiakoumou2008}.

\subsection*{Imaging without scattering}

Prior to imaging through turbid samples, the performance of the microscope was tested without scattering. Fig.~\ref{fig:No_Scattering} shows the images of the fluorescent beads and HEK cells obtained with different CR. Image quality in the case of beads remains practically the same both visually and in terms of signal-to-background ratio (SBR) for all levels of compression. High CR could not be achieved for cell imaging mainly because images are not sparse enough.

\subsection*{Lateral resolution through scattering}
According to Nyquist's sampling criterion applied to imaging, in a standard microscope a small detail of a specimen has to be sampled by at least two pixels on a camera in order to obtain a well-resolved image. Therefore, we define the lateral resolution in TRAFIX as the size of two pixels in the reconstructed images. 

A fluorescent micropattern was imaged through 490 $\upmu$m of scattering phantom with two different sets of Hadamard bases to show the increase in resolution when using a larger number of patterns (Fig.~\ref{fig:Dog}). With a FOV of approximately 90$\times$90 $\upmu$m$^2$ and using a 32$\times$32 pixel basis (1024 patterns), the lateral resolution is 5.6 $\upmu$m; and using a 64$\times$64 pixel basis (4096 patterns) it becomes 2.8 $\upmu$m. The resolution can also be improved by making the FOV smaller using higher magnification optics.

\subsection*{Imaging volumetric fluorescent samples}

A scattering phantom with fluorophores extending its entire volume was imaged to determine the performance of TRAFIX in imaging in presence of intense out-of-focus fluorescent light. The phantom contained 0.4 $\mum$ and 4.8 $\mum$ diameter fluorescent beads, and 1 $\mum$ diameter polystyrene beads embedded in 1.5 \% agarose. The 0.4 $\mum$ fluorescent microparticles contribute to creating a fluorescent background and the 4.8 $\mum$ beads simulate features of interest. This experiment was done with a Nikon 20$\times$ NA = 0.75 air objective. 

Fig.~\ref{fig:3Dsample_beads} shows an image of the phantom at a depth of $\sim$300 $\mum$. We remark that no reference images are shown because this is a three-dimensional sample and the imaged plane is not accessible from the reference imaging system. As the axial confinement of the TF illumination plane with the present objective at a depth of $\sim$300 $\mum$ is $\sim$20 $\mum$ (See section S4), 4.8 $\mum$ beads out of the focal plane generate a strong fluorescent background. However, beads located at the focal plane can still be clearly distinguished by simply adjusting the gray levels of the image. Axial resolution can be enhanced by changing imaging parameters such as magnification, NA or pulse duration (36), or by relaying on line-scanning TF illumination instead of wide-field.

\subsection*{Acquisition time}

The total acquisition time of TRAFIX can be estimated with the equation $T=2n^2(t_{exp}+t_{SLM})$ given in the manuscript. For example, a typical image through a scattering medium such as Fig. \ref{fig:SBR_smiley_colon} (d), was obtained using a full basis of 1024 patterns (32$\times$32 pixels) with an exposure time of 0.1 s per pattern and 0.03 s refresh time of the spatial light modulator (SLM). Therefore, the total acquisition time was approximately $T=2\cdot 1024\cdot(0.1+0.03)= 266 s \approx 5$ min. This time is increased when imaging deeper in biological tissue or using dimmer fluorescent samples because the exposure time of the camera has to be increased. Increasing the total number of patterns to achieve higher resolution also increases acquisition time. The longest duration experiment (Fig. 3 (d)) took approximately 1 hour because of the high resolution (64$\times$64 pixels) and of the low cellular fluorescence (HEK cells). 

In this present embodiment of TRAFIX, the acquisition speed is mainly limited by the low sensitivity and readout speed of the EMCCD camera and the slow refresh rate of the SLM ($\sim$30 Hz real-time rate). To speed up the imaging, the EMCCD camera would be replaced with a fast, sensitive photodetector such as a photomultiplier tube (PMT), often used for fast imaging in standard point-scanning two-photon microscopy (2PM). 2PM using a PMT routinely achieves frame rates of 15 Hz for images containing 1M pixels (e.g. used on a Nikon A1R HD microscope). The pixel dwell time is thus approximately 64 ns/pixel. In such a case, the imaging speed would no longer be limited by the exposure time of the detector. The next issue for improving acquisition time is the SLM. As TRAFIX currently uses binary Hadamard patterns, the SLM may be replaced with a significantly faster digital micromirror device (DMD, estimated to run at 22 kHz). In this way, the acquisition speed could be dramatically increased to real-time frame rate imaging. Using the noted equation, it can be shown that a state-of-the-art DMD could achieve frame rates of $\sim$10 Hz for a resolution of 32$\times$32 pixels, $\sim$3 Hz for 64$\times$64 pixels and $\sim$0.7 Hz for 128$\times$128 pixels assuming measurements with full bases. 

Imaging speed can be further increased by relying on compressive sensing. The effectiveness of compressive sensing is determined by the sparsity in the datasets, a high resolution image can in general be compressed to a higher extent than a low resolution one \cite{Sun2017}. Therefore, increasing the size of the pattern basis makes it possible to achieve a higher compression. Compression of 1\% and even higher has consistently been reported in the literature \cite{Olivas2013} suggesting that frame rates for 128$\times$128 pixels images can easily be increased to over $\sim$30 Hz~\cite{Higham2018,Xu2018}, enabling studies in time varying turbulence~\cite{Edgar2015}. 

Image reconstruction in the current study is performed offline on a computer with an Intel Core i7-6700K 4GHz processor and a RAM of 32 GB. The computer processing time depends on the level of sophistication of the algorithm. For example, an image can be reconstructed in 5 s by simply using eq. \ref{eq_1}. If noise reduction or compressed reconstruction is needed, longer processing times are typically required. The image reconstruction time using an orthogonal matching pursuit algorithm takes about 30 s for 32$\times$32 pixel images and 
it scales with the number of patterns used in the image reconstruction.  This can be reduced by use of new algorithms to achieve fast real-time reconstruction at video rate~\cite{Higham2018}.




\section{Comparison between TRAFIX and point-scanning two-photon microscopy (2PM)}\label{supp-TRAFIXvs2PM}

\subsection*{Signal-to-background ratio (SBR) at depth}

To compare the performance of TRAFIX and 2PM in imaging at depth, two experiments were carried out. They consisted in imaging fluorescent micropatterns at different depths through a scattering phantom ($l_{s}\approx100\mum$) and comparing the signal-to-background ratio (SBR) of the acquired images.

An underfilled Nikon 20$\times$ NA = 0.75 air objective was used for the studies below to equate lateral and axial resolution for both approaches.

- Experiment 1: comparable lateral resolution.  Firstly, the back aperture of the objective was filled so as to produce a focused spot of 3.75 $\mum$ diameter to make it similar to the lateral resolution of TRAFIX  (FOV of 84$\times$84 $\mum^2$ using an illumination basis size of 1024). We term this 2PM-Lateral.

- Experiment 2:  comparable axial resolution. Here, the NA of the illumination beam was increased to improve the axial resolution of 2PM to 8.4 $\mum$, which was the same as for TRAFIX (Fig. \ref{fig:axial} (c)). We term this 2-PM-Axial. 

The laser power was set to the maximum for TRAFIX, obtaining a power per unit area and pattern on the sample of 0.068 mW$/{\mum^2\cdot}$pattern. The total accumulated power per unit area after a measurement with the full basis was 69 mW$/\mum^2$. The laser power per unit area was then adjusted to generate approximately the same fluorescence intensity in 2PM as the total fluorescence intensity generated by TRAFIX after 1024 patterned illuminations. This was 2.2 mW$/\mum^2$. The exposure time of the camera was set to the same value for each pixel in 2PM scan as for each TRAFIX pattern to make the experiment fully comparable. As expected, in the experiments without scattering, the SBR was equivalent for both TRAFIX and 2PM (Fig. \ref{fig:SBR_smiley_depth}). All images were obtained with different micropatterns (smiley faces) to avoid the effect of photobleaching in the SBR calculation.

The results for both experiments are shown together in Fig. \ref{fig:SBR_smiley_depth}. We confirm that the SBR in 2PM degrades more rapidly than for TRAFIX when imaging at depth. Furthermore, we also performed these two experiments in exactly the same conditions through 200 $\mum$ of unfixed human colon tissue. Fig. \ref{fig:SBR_smiley_colon} also shows a higher SBR for TRAFIX at depth. These results show that TRAFIX achieves between 2 and 5 times higher SBR than 2PM for the samples and depths considered in this study. As described in the article, this is due to the resilience of TF beams to speckle formation and the fact that the temporal profile of femtosecond pulses is distorted to a lower extent compared to the spatial profile. 







\subsection*{Axial confinement}

The axial confinement of TRAFIX and 2PM was compared through different layers of a scattering phantom having a scattering mean free path of $l_s\approx 150 \mum$ and using a Nikon 20$\times$ NA = 0.75 objective.

Starting from the same axial confinement without scattering, Fig.~\ref{fig:axial} shows that TRAFIX (or equivalently, wide-field TF) degrades faster than 2PM. The axial extent of the illuminated plane in TRAFIX increases $\sim$2.6 times after $\sim$2.4 scattering mean-free-path lengths while in 2PM it only increases by $\sim$1.8 times. This behaviour for TF illumination has already been identified by H. Dana and S. Shoham (36).  It is also worth remarking that 2PM has better axial confinement than TRAFIX because it is currently based on wide-field TF illumination. To achieve the same axial confinement as 2PM, fast line-scanning TF illumination should be used to create the illumination patterns in TRAFIX~\cite{Durst2008}.


\subsection*{Photobleaching}

When imaging biological samples, photobleaching and photodamage caused by the high irradiation intensities of the illuminating light need to be carefully considered. The wide-field temporal focusing nature of TRAFIX avoids the need of spatially focusing a laser beam on the sample resulting in a reduced illumination power per unit area. 

The use of patterned illumination is also beneficial for imaging sensitive biological samples since the total illumination energy is distributed over many low intensity exposures. Even though the total accumulated power per unit area in TRAFIX may be higher than in 2PM, it has been shown that by delivering the same total light dosage in a sequential rather than instantaneous manner, photobleaching and photodamage are drastically reduced~\cite{Boudreau2016}. 

To verify this, we performed an experiment in which we sequentially imaged two fluorescent micropatterns without scattering with TRAFIX and with 2PM. The illumination power per unit area of TRAFIX was set to 0.034 mW/$\mum^2\cdot$pattern and the resulting accumulated power per unit area after projecting 1024 patterns was 35 mW/$\mum^2$. We determined that in order to have the same total fluorescence count (i.e. equivalent image quality) the laser power per unit area of 2PM in this case should be 1.1 mW/$\mum^2$. We observed that under this power per unit area, the fluorescent micropattern was severely damaged so we decided to perform this experiment with half of that power, i.e. 0.55 mW/$\mum^2$.

\begin{figure}[htbp]
\centering
\includegraphics[width=0.95\linewidth]{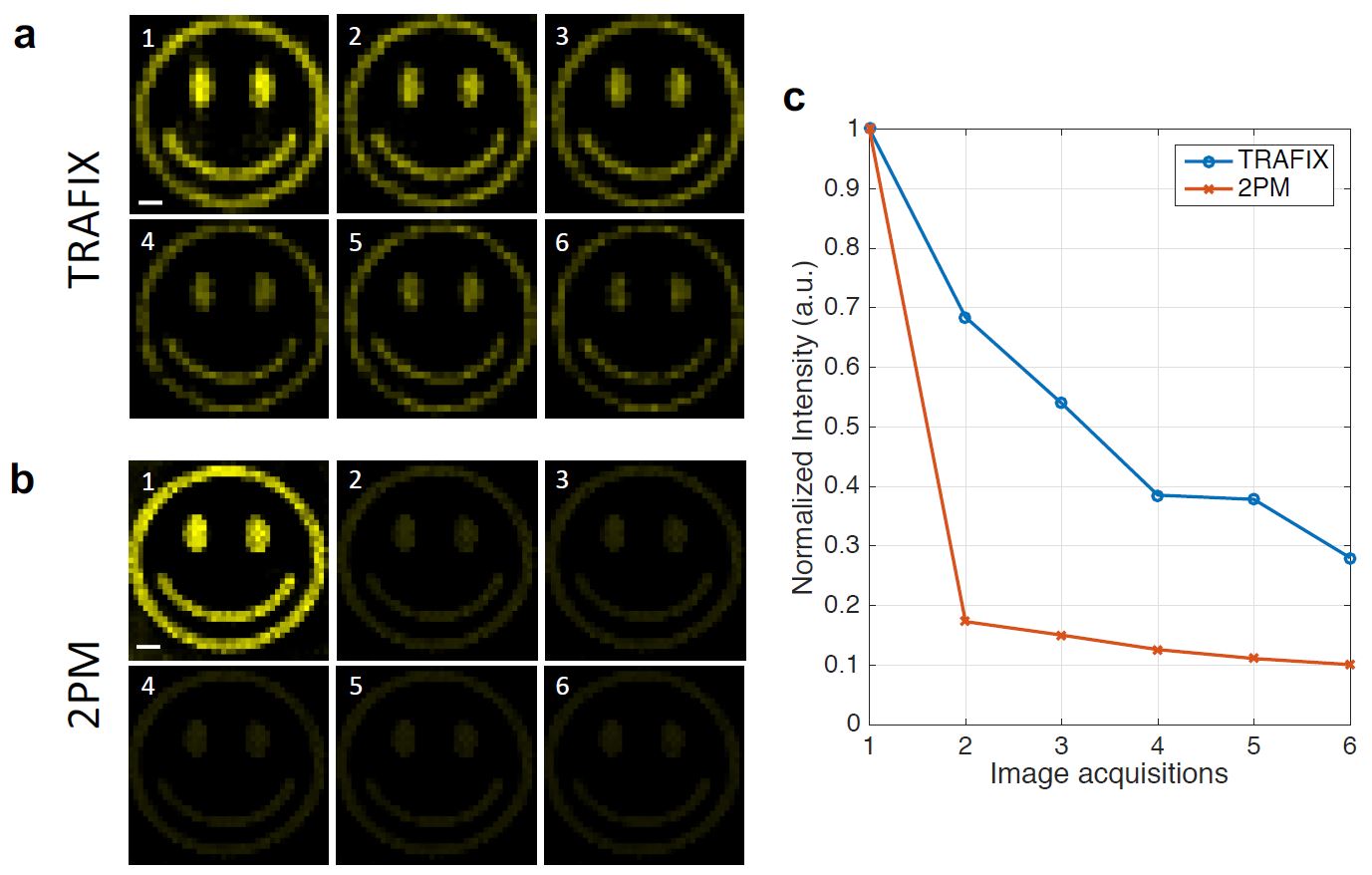}
\caption{\textbf{Photobleaching comparison of TRAFIX and point-scanning two-photon microscopy (2PM).} Successive images obtained with \textbf{(a)} TRAFIX and with \textbf{(b)} 2PM without scattering. Laser power per unit area for TRAFIX is 0.034 mW/$\mum^2\cdot$pattern and the resulting accumulated power per unit area after projecting 1024 patterns is 35 mW/$\mum^2$. The laser power per unit area for 2PM is 0.55 mW/$\mum^2$, which corresponds to half of the required power per unit area to generate the same total fluorescence signal as TRAFIX. Images taken with an underfilled 20$\times$ NA = 0.75 objective. \textbf{(c)} Fluorescence intensity decay after several acquisitions for TRAFIX and 2PM.}
\label{fig:bleach}
\end{figure}

\begin{figure}[htbp]
\centering
\includegraphics[scale=0.7]{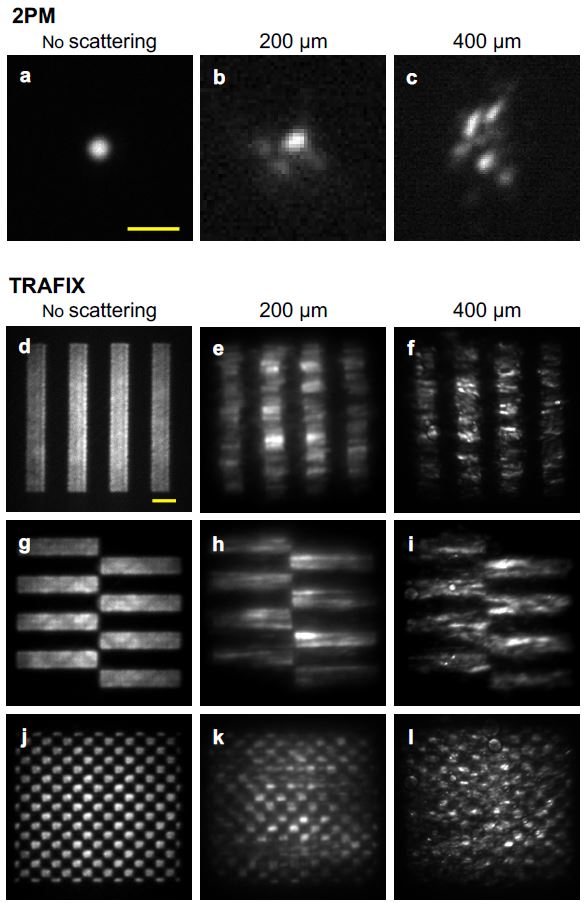} 
\caption{\textbf{Effect of scattering on illumination beams in point-scanning two-photon microscopy (2PM) and TRAFIX.} The three columns correspond to the illumination patterns without scattering, through 200 $\upmu$m and 400 $\upmu$m of unfixed colon tissue, respectively. (\textbf{a}-\textbf{c}) 4 $\mum$ diameter focused beam in 2PM. (\textbf{d} -\textbf{l}) Hadamard patterns of different orientation and spatial frequencies in TRAFIX. The size of the patterns is 130$\times$130$\mum^2$. Scale bars are 10 $\upmu$m in (\textbf{a}) and 20 $\upmu$m in (\textbf{d}). This experiment was done with a 40$\times$ NA = 0.8 objective.}
\label{fig:patterns}
\end{figure}

Fig.~\ref{fig:bleach} (a) and (b) show six images acquired sequentially with TRAFIX and 2PM, respectively. It is clear to see that the image quality of 2PM degrades much faster than that of TRAFIX even under lower laser power per unit area than required to obtain the same SBR. Fig.~\ref{fig:bleach} (c) shows the fluorescence intensity decay in both techniques. Under more favorable conditions compared to TRAFIX, 2PM photobleaching is still over three times higher. Thus, this highlights a further major advantage of the TRAFIX approach.



\subsection*{Effect of scattering on excitation beams}

A further experiment was performed to determine the effect of scattering on the illumination patterns of TRAFIX and 2PM.

A fluorescent layer placed on top of the colon tissue was illuminated with various TRAFIX patterns and the 2PM focused beam, and imaged from the reference arm. Fig. \ref{fig:patterns} shows the illumination patterns without scattering and through 200$\mum$ and 400$\mum$ of colon tissue. It can be seen that the 2PM focused spot turns into a speckle pattern after $\sim$400$\mum$ while low frequency Hadamard patterns remain relatively well conserved though with important brightness inhomogeneities.

\section{Polarisation state evaluation}\label{supp-polarization}

\begin{figure}[htbp]
\centering
\includegraphics[scale=0.8]{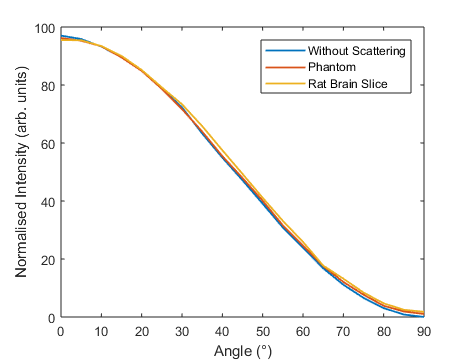}
\caption{\textbf{Effect of turbid media on light polarisation.}   
Linearly polarised illumination light retains its polarisation state through propagation in the scattering phantom or rat brain tissue used in this experiment. The data clearly obeys Malus' Law. The thickness of the scattering phantom and the rat brain tissue were 640 $\upmu$m and 400 $\upmu$m, respectively.}
\label{fig:polarization}
\end{figure}

A scattering phantom and rat brain tissue were illuminated with linearly polarised light, and the polarisation state of the transmitted light was analyzed using a linear polariser and a power meter. 
The degree of linear polarisation [$DOLP = (I_{\parallel}-I_{\bot})/(I_{\parallel}+I_{\bot})$] was measured in all three cases obtaining 0.999 for the incident light, and 0.978 and 0.964 for the transmitted light through the scattering phantom and rat brain tissue, respectively. 
Figure \ref{fig:polarization} shows that the measured data clearly obeys Malus' Law. In the figure, intensity was normalised taking into account the transmittance of the different samples and the polariser. 
We confirm that the polarisation state of light remains practically unchanged after propagating through the samples used in this investigation. Light only loses its polarisation state in samples that are multiple times thicker than the transport mean free path $l_t=l_s / (1-g)$, ($l_t>>l_s$ for most biological tissues) \cite{deAguiar2017}.


\clearpage
{\large\textbf{Supplementary Figures}}

\begin{table}[htbp]
\centering
\captionsetup{labelformat=empty}
\begin{tabular}{|c|p{12cm}|c|}
\hline
Table \# & Description & Page \# \\ \hline
    \ref{fig:tfbeams}     &     Numerically simulated TF laser beam propagating through 400 $\upmu$m of brain tissue.      &     ~\pageref{fig:tfbeams}    \\ \hline
    \ref{fig:tfsimulation}    &    Properties of a numerically simulated TF laser beam through brain tissue.     &     ~\pageref{fig:tfsimulation}      \\ \hline
    \ref{fig:Beam_Profile}    &     Effect of scattering on the beam profile with and without TF.        &     ~\pageref{fig:Beam_Profile}     \\ \hline
        \ref{fig:Depth_Profiles}    &     Depth profile of a TF beam through scattering phantom.        &     ~\pageref{fig:Depth_Profiles}     \\ \hline
        \ref{fig:FWHM_Fluor}    &     Characterisation of a TF beam through a scattering phantom.       &     ~\pageref{fig:FWHM_Fluor}     \\ \hline
        \ref{fig:No_Scattering}    &    Images of fluorescent microscopic samples without scattering.      &     ~\pageref{fig:No_Scattering}     \\ \hline
        \ref{fig:Dog}    &    Comparison of a hidden object and retrieved images through a scattering phantom with different resolution.    &     ~\pageref{fig:Dog}     \\ \hline
        \ref{fig:3Dsample_beads}    &    Image of 4.8 $\mum$ fluorescent beads in a volumetric scattering phantom.   &     ~\pageref{fig:3Dsample_beads}     \\ \hline
        \ref{fig:SBR_smiley_depth}    &    Comparison of signal-to-background ratio (SBR) of TRAFIX and point-scanning two-photon microscopy (2PM) at depth.   &     ~\pageref{fig:SBR_smiley_depth}     \\ \hline
        \ref{fig:SBR_smiley_colon}    &   Comparison of TRAFIX and point-scanning two-photon microscopy (2PM) through human colon tissue.    &     ~\pageref{fig:SBR_smiley_colon}     \\ \hline
        \ref{fig:axial}    &    Axial confinement in TRAFIX and point-scanning two-photon microscopy (2PM).    &     ~\pageref{fig:axial}     \\ \hline
        \ref{fig:bleach}    &    Photobleaching comparison of TRAFIX and point-scanning two-photon microscopy (2PM).    &     ~\pageref{fig:bleach}     \\ \hline
        \ref{fig:patterns}    &    Effect of scattering on illumination beams in point-scanning two-photon microscopy (2PM) and TRAFIX.    &     ~\pageref{fig:patterns}     \\ \hline
        \ref{fig:polarization}    &    Effect of turbid media on light polarisation.    &     ~\pageref{fig:polarization}     \\ \hline
\end{tabular}
\end{table}
\clearpage

\clearpage

\clearpage
{\large\textbf{Supplementary Tables}}

\begin{table}[htbp]
\centering
\captionsetup{labelformat=empty}
\begin{tabular}{|c|p{12cm}|c|}
\hline
Table \# & Description & Page \# \\ \hline
    \ref{SNR}     &     Signal-to-background ratio (SBR) measured for all the images shown in this work.        &     ~\pageref{SNR}    \\ \hline
    \ref{Table_Cells}    &    Cell diameters of all images shown in this work.      &     ~\pageref{Table_Cells}      \\ \hline
    \ref{Table_Beads}    &     Beads spacing corresponding to all images shown in this work.        &     ~\pageref{Table_Beads}     \\ \hline
\end{tabular}
\end{table}

\clearpage

\setlength{\arrayrulewidth}{0.3mm}
\setlength{\tabcolsep}{6pt}
\renewcommand{\arraystretch}{1.5}
\newcommand\VRule[1][\arrayrulewidth]{\vrule width #1}
\begin{table}[htbp]
\centering
\begin{tabular}{|c|c|c|c|c|c|}
\hline
 & \multicolumn{4}{c|}{\cellcolor[HTML]{EFEFEF}\textbf{SIGNAL-TO-BACKGROUND RATIO}} \\ \cline{2-5} 
\multirow{-2}{*}{} & \cellcolor[HTML]{B3E7B3}CS (CR=1) & \cellcolor[HTML]{B3E7B3}CS (CR=2) & \cellcolor[HTML]{B3E7B3}CS (CR=4) & \cellcolor[HTML]{B3E7B3}CS (CR=8) \\ \hline
\rowcolor[HTML]{EFEFEF} 
\cellcolor[HTML]{FFCCC9}Beads Without Scattering &14.9 $\pm$ 1.5 & 11.6 $\pm$ 1.0 & 6.4 $\pm$ 0.7 & 3.7 $\pm$ 0.4 \\ \hline
\cellcolor[HTML]{FFCCC9}Beads Phantom  &5.5 $\pm$ 2.1  &4.2 $\pm$ 1.8  &3.9 $\pm$ 1.5  &2.6 $\pm$ 0.7  \\ \hline
\cellcolor[HTML]{FFCCC9}Beads Colon  &5.4 $\pm$ 1.0  &-- &-- &--  \\ \hline
\rowcolor[HTML]{EFEFEF} 
\cellcolor[HTML]{FFCCC9}Cells Without Scattering  &14.8 $\pm$ 2.6  &11.5 $\pm$ 1.6  &--  &--  \\ \hline
\cellcolor[HTML]{FFCCC9}Cells Phantom  &4.9 $\pm$ 1.5  &4.3 $\pm$ 0.4  &--  &--  \\ \hline
\cellcolor[HTML]{FFCCC9}Cells Colon  &4.6 $\pm$ 0.3  &--  &--  &--  \\ \hline
\end{tabular}
\caption{\textbf{Signal-to-background ratio (SBR) measured for images shown in this work.} Numerical columns correspond to retrieved images obtained through scattering media with TRAFIX at different compression ratios (CR).}
\label{SNR}
\end{table}

\begin{table}[htbp]
\centering
\begin{tabular}{|c|c|c|c|c|c|l|}
\hline
\rowcolor[HTML]{C0C0C0} 
\cellcolor[HTML]{FFFFFF} & \cellcolor[HTML]{FFFFFF} & \multicolumn{3}{c|}{\cellcolor[HTML]{C0C0C0}\textbf{Diameter FWHM ($\upmu$m)}} & \multicolumn{2}{c|}{\cellcolor[HTML]{C0C0C0}\textbf{Deviation (\%)}} \\ \cline{3-7} 
\rowcolor[HTML]{B3E7B3} 
\multirow{-2}{*}{\cellcolor[HTML]{FFFFFF} } & \multirow{-2}{*}{\cellcolor[HTML]{FFFFFF}\textbf{Cell}} & \textbf{Reference} & \textbf{CR=1} & \textbf{CR=2} & \textbf{CR=1} & \textbf{CR=2} \\ \hline
\rowcolor[HTML]{FFFFFF} 
\cellcolor[HTML]{FFFFFF} & 1 & 19.20 & 18.92 & 20.44 & 1.5 & \multicolumn{1}{c|}{6.4} \\ \cline{2-7} 
\rowcolor[HTML]{FFFFFF} 
\multirow{-2}{*}{\cellcolor[HTML]{FFFFFF}\begin{tabular}[c]{@{}c@{}}Without\\ scattering\end{tabular}} & 2 & 16.31 & 16.35 & 14.73 & 0.2 & \multicolumn{1}{c|}{9.7} \\ \hline
\rowcolor[HTML]{EFEFEF} 
\cellcolor[HTML]{EFEFEF} & 1 & 20.66 & 20.64 & 19.23 & 0.1 & \multicolumn{1}{c|}{\cellcolor[HTML]{EFEFEF}6.9} \\ \cline{2-7} 
\rowcolor[HTML]{EFEFEF} 
\multirow{-2}{*}{\cellcolor[HTML]{EFEFEF}\begin{tabular}[c]{@{}c@{}}Scattering\\ phantom\end{tabular}} & 2 & 14.32 & 13.86 & 12.71 & 3.3 & \multicolumn{1}{c|}{\cellcolor[HTML]{EFEFEF}11.2} \\ \hline
\rowcolor[HTML]{FDE1DF} 
 & 1 & 13.07 & 13.91 &-- & 6.5 &\multicolumn{1}{c|}{--} \\ \cline{2-7} 
 \rowcolor[HTML]{FDE1DF} 
 & 2 & 13.94 & 11.71 &-- & 16.0 &\multicolumn{1}{c|}{--} \\ \cline{2-7} 
  \rowcolor[HTML]{FDE1DF} 
 & 3 & 13.22 & 12.07 &-- & 8.6 &\multicolumn{1}{c|}{--} \\ \cline{2-7} 
  \rowcolor[HTML]{FDE1DF} 
 & 4 & 12.22 & 13.32 &-- & 9.1 &\multicolumn{1}{c|}{--} \\ \cline{2-7} 
\rowcolor[HTML]{FDE1DF} 
\multirow{-5}{*}{\begin{tabular}[c]{@{}c@{}}Colon\\ tissue\end{tabular}} & 5 & 11.65 & 12.53 & -- & 7.6 & \multicolumn{1}{c|}{--} \\ \hline
\end{tabular}
\caption{\textbf{Cell diameters of images shown in this work.} Cells correspond, in order, to Fig.~\ref{fig:No_Scattering}, Fig.~\ref{fig:Phantom} and Fig.~\ref{fig:Colon}. The deviation from the reference value is shown in the last two columns.}
\label{Table_Cells}
\end{table}

\clearpage
\begin{table}[htbp]
\centering
\begin{tabular}{|c|c|c|c|c|c|c|c|c|c|c|}
\hline
\multicolumn{1}{|l|}{} & \cellcolor[HTML]{FFFFFF} & \multicolumn{5}{c|}{\cellcolor[HTML]{C0C0C0}\textbf{Distance ($\upmu$m)}} & \multicolumn{4}{c|}{\cellcolor[HTML]{C0C0C0}\textbf{Deviation (\%)}} \\ \cline{3-11} 
\multicolumn{1}{|l|}{\multirow{-2}{*}{}} & \multirow{-2}{*}{\cellcolor[HTML]{FFFFFF}\textbf{No.}} & \cellcolor[HTML]{B3E7B3}\textbf{Ref.} & \cellcolor[HTML]{B3E7B3}\textbf{CR=1} & \cellcolor[HTML]{B3E7B3}\textbf{CR=2} & \cellcolor[HTML]{B3E7B3}\textbf{CR=4} & \multicolumn{1}{l|}{\cellcolor[HTML]{B3E7B3}\textbf{CR=8}} & \multicolumn{1}{l|}{\cellcolor[HTML]{B3E7B3}\textbf{CR=1}} & \multicolumn{1}{l|}{\cellcolor[HTML]{B3E7B3}\textbf{CR=2}} & \multicolumn{1}{l|}{\cellcolor[HTML]{B3E7B3}\textbf{CR=4}} & \multicolumn{1}{l|}{\cellcolor[HTML]{B3E7B3}\textbf{CR=8}} \\ \hline
\multicolumn{1}{|l|}{} & \cellcolor[HTML]{FFFFFF}1 & \cellcolor[HTML]{FFFFFF}42.19 & \cellcolor[HTML]{FFFFFF}40.99 & \cellcolor[HTML]{FFFFFF}40.29 & \cellcolor[HTML]{FFFFFF}41.46 & \cellcolor[HTML]{FFFFFF}40.38 & \cellcolor[HTML]{FFFFFF}2.8 & \cellcolor[HTML]{FFFFFF}4.5 & \cellcolor[HTML]{FFFFFF}1.7 & \cellcolor[HTML]{FFFFFF}4.2 \\ \cline{2-11} 
\multicolumn{1}{|l|}{} & \cellcolor[HTML]{FFFFFF}2 & \cellcolor[HTML]{FFFFFF}20.40 & \cellcolor[HTML]{FFFFFF}19.16 & \cellcolor[HTML]{FFFFFF}19.81 & \cellcolor[HTML]{FFFFFF}19.73 & \cellcolor[HTML]{FFFFFF}19.54 & \cellcolor[HTML]{FFFFFF}6.1 & \cellcolor[HTML]{FFFFFF}2.9 & \cellcolor[HTML]{FFFFFF}3.3 & \cellcolor[HTML]{FFFFFF}4.3 \\ \cline{2-11} 
\multicolumn{1}{|l|}{\multirow{-3}{*}{\begin{tabular}[c]{@{}l@{}}Without \\ scattering\end{tabular}}} & \cellcolor[HTML]{FFFFFF}3 & \cellcolor[HTML]{FFFFFF}55.25 & \cellcolor[HTML]{FFFFFF}52.95 & \cellcolor[HTML]{FFFFFF}52.15 & \cellcolor[HTML]{FFFFFF}54.02 & \cellcolor[HTML]{FFFFFF}52.88 & \cellcolor[HTML]{FFFFFF}4.2 & \cellcolor[HTML]{FFFFFF}5.6 & \cellcolor[HTML]{FFFFFF}2.2 & \cellcolor[HTML]{FFFFFF}4.2 \\ \hline
\rowcolor[HTML]{EFEFEF} 
\cellcolor[HTML]{EFEFEF} & 1 & 23.21 & 23.07 & 23.05 & 23.51 & 23.57 & 0.6 & 0.7 & 1.3 & 1.6 \\ \cline{2-11} 
\rowcolor[HTML]{EFEFEF} 
\cellcolor[HTML]{EFEFEF} & 2 & 18.36 & 17.87 & 18.04 & 18.19 & 18.80 & 2.7 & 1.7 & 0.9 & 2.4 \\ \cline{2-11} 
\rowcolor[HTML]{EFEFEF} 
\multirow{-3}{*}{\cellcolor[HTML]{EFEFEF}\begin{tabular}[c]{@{}c@{}}Scattering\\ phantom\end{tabular}} & 3 & 39.90 & 40.04 & 40.04 & 39.37 & 39.69 & 0.4 & 0.4 & 1.3 & 0.5 \\ \hline
\rowcolor[HTML]{FDE1DF} 
\cellcolor[HTML]{FDE1DF} & 1 & 13.38 & 15.39 & - & - & - & 15.0 & - & - & - \\ \cline{2-11} 
\rowcolor[HTML]{FDE1DF} 
\cellcolor[HTML]{FDE1DF} & 2 & 38.31 & 44.55 & - & - & - & 16.3 & - & - & - \\ \cline{2-11} 
\rowcolor[HTML]{FDE1DF} 
\multirow{-3}{*}{\cellcolor[HTML]{FDE1DF}\begin{tabular}[c]{@{}c@{}}Colon\\ tissue\end{tabular}} & 3 & 30.77 & 35.88 & - & - & - & 16.0 & - & - & - \\ \hline
\end{tabular}
\caption{\textbf{Beads spacing corresponding to images shown in this work.} They refer, in order, to Fig.~\ref{fig:No_Scattering}, Fig.~\ref{fig:Phantom} and Fig.~\ref{fig:Colon}. The deviation from the reference value is shown in the last four columns.}
\label{Table_Beads}
\end{table}

\end{document}